\documentclass[12pt,psamsfonts]{article}
\usepackage{amsmath,amssymb,amsthm}
\usepackage{eucal}
\usepackage[all,cmtip]{xy}
\usepackage{bm}
\usepackage{hyperref}

\numberwithin{equation}{section}

\usepackage{geometry}
\def\a{\alpha}

\def\g{\gamma}
\def\d{\delta}
\def\e{\epsilon}

\def\th{\theta}

\def\l{\lambda}

\def\s{\sigma}

\def\G{\Gamma}

\def\L{\Lambda}

\def\Om{\Omega}

\def\RR{\mathbb R}

\newcommand{\C}[1]{$(\ref{#1})$}
\newcommand{\hph}[1]{{\hphantom{#1}}}
\newcommand{\ov}[1]{{\overline{#1}}}
\def\o{\over}
\def\pa{\partial}

\def\hlf{\frac{1}{2}}
\def\lp{\left(}
\def\rp{\right)}
\def\ls{\left[}
\def\rs{\right]}

\def\R{{\Bbb R}}

\def\lp{\left(}
\def\rp{\right)}
\def\ls{\left[}
\def\rs{\right]}

\def\algg{\mathfrak g}

\newcommand{\be}{\begin{equation}}
\newcommand{\ee}{\end{equation}}
\newcommand{\bea}{\begin{eqnarray}}
\newcommand{\eea}{\end{eqnarray}}

\newcommand{\La}{\Lambda}
\newcommand{\la}{\lambda}

\newcommand{\al}{\alpha}
\newcommand{\w}{\wedge}
\newcommand{\Hom}{\operatorname{Hom}}

\newcommand{\mcA}{\mathcal{A}}

\newcommand{\mcD}{\mathcal{D}}
\newcommand{\mcF}{\mathcal{F}}

\newcommand{\mcW}{\mathcal{W}}

\newcommand{\non}{\nonumber}

\makeatletter
\renewcommand\section{\@startsection {section}{1}{\z@}%
                                   {-3.5ex \@plus -1ex \@minus -.2ex}
                                   {2.3ex \@plus.2ex}%
                                   {\normalfont\large\bfseries}}
\renewcommand\subsection{\@startsection{subsection}{2}{\z@}%
                                     {-3.25ex\@plus -1ex \@minus -.2ex}%
                                     {1.5ex \@plus .2ex}%
                                     {\normalfont\bfseries}}
\makeatother

\begin{document}

\begin{center}
\addtolength{\baselineskip}{.5mm}
\thispagestyle{empty}
\begin{flushright}
{\sc MI-TH-1612}\\
\end{flushright}

\vspace{20mm}

{\Large  \bf 
Chern-Simons Actions and Their Gaugings\\  
in 4D, $\bm {N=1}$ Superspace
}
\\[15mm]
{Katrin Becker, Melanie Becker, William D. Linch III, and Daniel Robbins}
\\[5mm]
{\it George P. and Cynthia Woods Mitchell Institute for }\\
{\it Fundamental Physics and Astronomy, Texas A\& M University,}\\
{\it College Station, TX 77843-4242, USA}\\[5mm]

\vspace{10mm}

{\bf  Abstract}
\end{center}
We gauge the abelian hierarchy of tensor fields in 4D by a Lie algebra $\mathfrak{g}$.  The resulting non-abelian tensor hierarchy can be interpreted via a $\mathfrak{g}$-equivariant chain complex.  We lift this structure to $N=1$ superspace by constructing superfield analogs for the tensor fields, along with covariant superfield strengths.  Next we construct Chern-Simons actions, for both the bosonic and $N=1$ cases, and note that the condition of gauge invariance can be presented cohomologically.  Finally, we provide an explicit realization of these structures by dimensional reduction, for example by reducing the three-form of eleven-dimensional supergravity into a superspace with manifest 4D, $N=1$ supersymmetry.

\newpage
\thispagestyle{empty}
\tableofcontents

\newpage
\setcounter{page}1

\section{Introduction}

Gravitational tensor hierarchies are a common feature of supergravity compactifications resulting from the reduction of component $p$-forms in the higher-dimensional component spectrum that are charged under the higher-dimensional superdiffeomorphisms \cite{deWit:2005hv,deWit:2008ta,deWit:2008gc}. 
Upon compactification, some of the components of the gravitino generally become massive but leave behind massless non-abelian gauge fields from mixed components of the frame and their superpartners. 
What remains is a hierarchy of differential forms of various spacetime degrees, all charged under the residual diffeomorphisms compatible with the splitting of the compactified spacetime.
Further decoupling this structure from the lower-dimensional supergravity fields, one is left with a hierarchy of $p$-forms charged under the (non-abelian) gauge algebra of diffeomorphisms of the internal manifold. 

This gauged $p$-form hierarchy may be abstracted away from its gravitational avatar by replacing the algebra of diffeomorphisms with an arbitrary Lie algebra $\mathfrak g$ and assigning to each gauge $p$-form a representation $\rho_p: \mathfrak g \to GL(V_p)$. Consistency of the resulting ``non-abelian tensor hierarchy'' requires a complicated set of identities to hold between the Lie algebra, its representations, and a collection of maps relating the forms of various degrees \cite{Samtleben:2011fj,Samtleben:2012mi}. Attempts at interpreting this structure algebro-geometrically suggest that they are strongly homotopy Lie algebras \cite{Saemann:2012uq, Palmer:2013pka}.

Here, we take a somewhat different approach more closely related to the gravitational tensor hierarchy \cite{deWit:2009zv, Bergshoeff:2009ph,Hartong:2009az,Greitz:2013pua,Howe:2015hpa} in which the conditions on the couplings of the theory come from two requirements. 
The first set of conditions results from closure of the gauge algebra induced on the tower of $p$-forms by the representations $\rho_p$. Roughly speaking, this set says that the induced action of the gauge algebra on the tower of forms is $\mathfrak g$-covariant. 
The second set of conditions comes from requiring the existence of gauge-covariant field strengths for all fields in the tower. This defines the tower as a differential complex and defines an extension of the Lie derivative (naturally defined on $p$-foms) to this gauged complex. 
We refer to these two sets of conditions as the hierarchy equations.  Taken together, our gravitationally-motivated version of the non-abelian tensor hierarchy is a $\mathfrak g$-equivariant double complex constructed from de Rham forms with values in a complex of representations $\rho_p$ of $\mathfrak g$.

For applications to supergravity and the construction of superconformal models, it is of interest to supersymmetrize the bosonic hierarchy. 
This hierarchy simplifies dramatically if we turn off the $\mathfrak g$ gauging and in reference \cite{Becker:2016xgv}, we embedded this ``abelian tensor hierarchy'' into 4D, $N=1$ superspace. 
In this paper we gauge this superspace hierarchy to obtain a non-abelian tensor hierarchy in 4D, $N=1$ superspace. 
We begin in section \ref{boshie} by coupling a system of bosonic $p$-forms to a non-abelian gauge field. The set of fields and their interactions are inspired by but not identical to the fields 
obtained from a Kaluza-Klein reduction of the three-form and a metric gauge field of eleven-dimensional supergravity to four dimension. In section \ref{abs0} we phrase the hierarchy equations
in the language of homological algebra. 
(We consider the abstract formulation important because it illuminates the meaning of the hierarchy equations and it gives hints about possible generalizations.)
In subsection \ref{intprod} we write the hierarchy equations in terms of Lie derivatives and interior products and we recover some familiar equations such as Cartan's magic formula. 
Then in section \ref{susyhie} we formulate this system in superspace, thereby gauging the abelian superspace hierarchy of reference \cite{Becker:2016xgv}. To set up the conventions and 
quote some results which are useful for the rest of the paper we recall how to formulate non-abelian gauge fields in superspace in section \ref{nonabsu}. Then in section \ref{inchie} we embed the bosonic fields and transformations into superfields. Moreover, we define field strengths and show that they transform covariantly.

Once this is done, we turn to the construction of Chern-Simons-like invariants, first in the bosonic case (section \ref{sec:BCSAs}) and then in superspace (section \ref{sec:SCSAs}) (previous approaches to supersymmetric Chern-Simons invariants include \cite{Howe:1998tsa,Berkovits:2008qw}). 
These constructions require the definition of certain cocycles on the tensor algebra of the total complex. Their (co)homological interpretation is relatively straightforward but explicit checking of their compatibility with the structure of the gauged hierarchy is somewhat involved, requiring repeated use of the hierarchy equations and superspace $D$-algebra identities. 
To illustrate the formalism and to show that the resulting structure admits non-trivial solutions, we turn in section \ref{sec:DimRed} to the explicit example of the Chern-Simons form of eleven-dimensional supergravity. Decomposition of the eleven-dimensional 3-form and its Chern-Simons 11-form into four-dimensional representations gives an explicit solution to the hierarchy equations and the required Chern-Simons cocycle conditions. Substitution into the superspace Chern-Simons action gives an embedding with manifest 4D, $N=1$ supersymmetry. 
We conclude in section \ref{sec:Prospects} with a summary of our result and comment on its relationship to related approaches and applications.

\section{Bosonic Hierarchy}\label{boshie}

Consider a collection of real scalars, one-forms, two-forms, and so on in $d$ space-time dimensions, $\phi^{I_p}_{[p]}$. The components of these forms are denoted by
\begin{equation}
\phi^{I_p}_{a_1 \dots a_p},
\end{equation}
and are functions taking values in real vector
spaces $V_p$. Here $I_p=1,\dots, \mathrm{dim}(V_p)$ labels the coordinates in some basis of
$V_p$.
These vector spaces are not necessarily finite dimensional. Space-time indices are labelled by lower case letters from the beginning of the Latin alphabet.
The fields considered herein are elements of
$\Om^q(\RR^d)\otimes V_p$ or $\Om^q(\RR^d)\otimes \mathfrak g$, for some $p$ and $q$. Here $\Om^\bullet(\RR^d)$ is the $d$-dimensional de-Rham complex and $\mathfrak g$ is a Lie algebra. In equations without explicit space-time indices, we use a subscript $[p]$ to indicate that the given object is a $p$-form.

There is a non-abelian gauge field ${\cal A}$ with transformation
\begin{equation}
\d {\cal A}^k_a =\pa_a \l^k+{f^k}_{lm} \l^l {\cal A}_a^m,
\end{equation}
and field strength
\begin{equation}
{\cal F}^k_{ab} = 2 \pa_{[a}{\cal A}^k_{b]} - {f^k}_{lm}{\cal A}^l_{[a}{\cal A}_{b]}^m.
\end{equation}
Here ${f^k}_{lm}$ are the structure constants of the gauge algebra. We have expanded ${\cal A}_a = {\cal A}_a^k T_k$, where $T_k$ are the generators of the gauge algebra.
Closure of the gauge algebra implies that the structure constants are anti-symmetric in their lower indices and the Jacobi identity holds,
\begin{equation}\label{eq:fClosure}
{f^k}_{(lm)}=0, \qquad {f^k}_{p[l} {f^p}_{mn]}=0.
\end{equation}
As a result the gauge algebra is a Lie algebra which is denoted by $\mathfrak g$.

For each $p>0$ there is a gauge transformation parameterized by a differential $(p-1)$-form $\L ^{I_p}_{[p-1]}$, which generates abelian $p$-form transformations. In addition, there is a shift by the parameter $\L ^{I_{p+1}}_{[p]}$
\begin{equation}
\d\phi^{I_p}_{a_1\cdots a_p}=p\pa _{[a_1}\L ^{I_p}_{a_2\cdots a_p]}+\lp q^{(p)}\rp^{I_p}_{\hph{I_p}J_{p+1}}\L ^{J_{p+1}}_{a_1\cdots a_p},
\end{equation}
where $(q^{(p)})^{I_p}_{\hph{I_p}J_{p+1}}$ are linear maps
\begin{equation}
q^{(p)}: V_{p+1} \to V_{p}.
\end{equation}
In the following we suppress the index $(p)$ on $q^{(p)}$ and write only $q$, whenever this index is clear from the context.

The tensor fields are charged under the non-abelian gauge transformation.
When coupled to the non-abelian gauge field the change of the tensor fields after infinitesimal gauge transformations is\footnote{This is not necessarily the most general possible form of the transformation, but it is sufficiently general to encompass the cases which arise from dimensional reduction.}
\begin{equation}\label{e27}
\begin{split}
\d\phi^{I_p}_{a_1\cdots a_p} & = \lp t_k\rp^{I_p}_{\hph{I_p}J_p}\l^k\phi^{J_p}_{a_1\cdots a_p}+p\pa_{[a_1}\L^{I_p}_{a_2\cdots a_p]}-p\lp t_k\rp^{I_p}_{\hph{I_p}J_p}{\cal A}^k_{[a_1}\L^{J_p}_{a_2\cdots a_p]}+q^{I_p}_{\hph{I_p}J_{p+1}}\L^{J_{p+1}}_{a_1\cdots a_p}\\
&  +\frac{p\lp p-1\rp}{2}\lp h_k\rp^{I_p}_{\hph{I_p}J_{p-1}}{\cal F}^k_{[a_1a_2}\L^{J_{p-1}}_{a_3\cdots a_p]},
\end{split}
\end{equation}
Here $t_k$ are a set of linear maps
\begin{equation}
t: (\Om^p \otimes \mathfrak g) \times (\Om^q \otimes V_r) \to \Om^{p+q} \otimes V_r,
\end{equation}
which very explicitly take
\begin{equation}
x^k_{a_1\cdots a_p}\in\Om^p\otimes\mathfrak{g},\qquad\varphi^{I_r}_{b_1\cdots b_q}\in\Om^q\otimes V_r,
\end{equation}
to
\begin{equation}
t(x,\varphi)^{I_r}_{a_1\cdots a_{p+q}}=\frac{\lp p+q\rp!}{p!q!}\lp t_k\rp^{I_r}_{\hph{I_r}J_r}x^k_{[a_1\cdots a_p}\varphi^{J_r}_{a_{p+1}\cdots a_{p+q}]}.
\end{equation}
While $h_k$ are linear maps
\begin{equation}
\label{eq:h}
h:\lp\Om^p\otimes\mathfrak{g}\rp\times\lp\Om^q\otimes V_r\rp\longrightarrow\Om^{p+q}\otimes V_{r+1},
\end{equation}
which act via
\begin{equation}
h(x,\varphi)^{I_{r+1}}_{a_1\cdots a_{p+q}}=\frac{\lp p+q\rp!}{p!q!}\lp h_k\rp^{I_{r+1}}_{\hph{I_{r+1}}J_r}x^k_{[a_1\cdots a_p}\varphi^{J_r}_{a_{p+1}\cdots a_{p+q}]}.
\end{equation}
Moreover, the maps $q$ have been trivially extended to
\begin{equation}
q:\Om^p\otimes V_q\longrightarrow\Om^p\otimes V_{q-1},
\end{equation}
by acting with the identity on the first factor.

Closure of the gauge algebra, {\it i.e.} requiring that the commutator of two transformations $\d, \d'$ of the type (\ref{e27}) gives another one
\begin{equation}
[\d, \d']= \d'',
\end{equation}
for some $\d''$,
requires
\begin{subequations}
\begin{align}
\label{e214a}
0=& \lp t_k\rp^{I_p}_{\hph{I_p}K_p}\lp t_l \rp^{K_p}_{\hph{K_p}J_p}-\lp t_l \rp^{I_p}_{\hph{I_p}K_p}\lp t_k\rp^{K_p}_{\hph{K_p}J_p}-f^m_{\hph{m}kl }\lp t_m\rp^{I_p}_{\hph{I_p}J_p},\\
\label{e214b}
0=& q^{I_p}_{\hph{I_p}K_{p+1}}\lp t_k\rp^{K_{p+1}}_{\hph{K_{p+1}}J_{p+1}}-\lp t_k\rp^{I_p}_{\hph{I_p}K_p}q^{K_p}_{\hph{K_p}J_{p+1}},\\
\label{e214c}
0=& \lp h_k\rp^{I_p}_{\hph{I_p}K_{p-1}}\lp t_l \rp^{K_{p-1}}_{\hph{K_{p-1}}J_{p-1}}-\lp t_l \rp^{I_p}_{\hph{I_p}K_p}\lp h_k\rp^{K_p}_{\hph{K_p}J_{p-1}}-f^m_{\hph{m}kl }\lp h_m\rp^{I_p}_{\hph{I_p}J_{p-1}}.
\end{align}
\end{subequations}
Equation (\ref{e214a}) says that the $t_i$ form a representation of the gauge algebra $\mathfrak g$. This action of the gauge algebra on the forms commutes with the map $q$ by (\ref{e214b}). Equation (\ref{e214c}) says that the pairing of gauge forms with ``matter'' forms defined by the $h$'s is covariant.

Field strengths are given by
\begin{equation}
\begin{split}
F^{I_p}_{a_1\cdots a_{p+1}} & = \lp p+1\rp\pa_{[a_1}\phi^{I_p}_{a_2\cdots a_{p+1}]}-\lp p+1\rp\lp t_k\rp^{I_p}_{\hph{I_p}J_p}{\cal A} ^k_{[a_1}\phi^{J_p}_{a_2\cdots a_{p+1}]}-q^{I_p}_{\hph{I_p}J_{p+1}}\phi^{J_{p+1}}_{a_1\cdots a_{p+1}}\\
& -\frac{p\lp p+1\rp}{2}\lp h_k\rp^{I_p}_{\hph{I_p}J_{p-1}}{\cal F} ^k_{[a_1a_2}\phi^{J_{p-1}}_{a_3\cdots a_{p+1}]}.
\end{split}
\end{equation}
These are covariant, {\it i.e.}
\begin{equation}
\d F^{I_p}_{a_1\cdots a_{p+1}}=\lp t_k\rp^{I_p}_{\hph{I_p}J_p}\l ^kF^{J_p}_{a_1\cdots a_{p+1}},
\end{equation}
provided that we also have
\begin{subequations}
\begin{align}\label{e217a}
0=& q^{I_p}_{\hph{I_p}K_{p+1}}q^{K_{p+1}}_{\hph{K_{p+1}}J_{p+2}},\\
\label{e217b}
0=& q^{I_p}_{\hph{I_p}K_{p+1}}\lp h_k\rp^{K_{p+1}}_{\hph{K_{p+1}}J_p}+\lp h_k\rp^{I_p}_{\hph{I_p}K_{p-1}}q^{K_{p-1}}_{\hph{K_{p-1}}J_p}+\lp t_k\rp^{I_p}_{\hph{I_p}J_p},\\
\label{e217c}
0=& \lp h_{k}\rp^{I_p}_{\hph{I_p}K_{p-1}}\lp h_{l }\rp^{K_{p-1}}_{\hph{K_{p-1}}J_{p-2}}+\lp h_{l}\rp^{I_p}_{\hph{I_p}K_{p-1}}\lp h_{k }\rp^{K_{p-1}}_{\hph{K_{p-1}}J_{p-2}}.
\end{align}
\end{subequations}

\section{Abstract Formulation} \label{abs0} 

In this section we recast the results of section 2 in the language of homological algebra. This simplifies the notation and suggests a natural interpretation of each of the hierarchy equations.

\subsection{Homological Algebra}\label{abs}

We consider a set of fields, field strengths, and gauge parameters which are sections of $\Om^q(\R^d)\otimes V_p$ or of $\Om^q(\R^d)\otimes\mathfrak{g}$ for some $p$ and $q$. We will drop the $\R^d$ below, but it should be considered implicit.  Specifically, we have table 1.
\begin{table}[!htp]
\centering
\begin{tabular}{|l|l|}
\hline
Object  & Bundle \\
\hline
$\l$ & $\Om^0\otimes\mathfrak{g}$ \\
${\cal A}$ & $\Om^1\otimes\mathfrak{g}$ \\
${\cal F} $ & $\Om^2\otimes\mathfrak{g}$ \\
\hline
$\L $ & $\Om^{p-1}\otimes V_p$ \\
$\phi$ & $\Om^p\otimes V_p$ \\
$F$ & $\Om^{p+1}\otimes V_p$ \\
\hline
\end{tabular}
\caption{Gauge parameters ($\l$, $\L$),
potentials (${\cal A}$, $\phi$) and field strengths (${\cal F}$, $F$), and the space each one lives in.  }
\end{table}

On these objects we also have the following operations. There is a set of linear operators
\begin{equation}
q:V_{p+1}\longrightarrow V_p,
\end{equation}
for each $p$, satisfying
\begin{equation}\label{e32}
q^2=0.
\end{equation}
The set of vector spaces $V_p$ can then be assembled into a chain complex $V_\bullet$,
\begin{equation}
\label{eq:VComplex}
V_\bullet: \cdots
\stackrel{q}{\rightarrow}V_{p+1}
\stackrel{q} {\rightarrow}V_p
\stackrel{q}{\rightarrow}V_{p-1}
\stackrel{q}{\rightarrow}\cdots
\stackrel{q}{\rightarrow}V_0.
\end{equation}
This can trivially be extended to a map
\begin{equation}
q:\Om^r\otimes V_{p+1}\longrightarrow\Om^r\otimes V_{p},
\end{equation}
by acting with the identity on the first factor.

The Lie bracket on $\mathfrak{g}$ is denoted by $[\cdot, \cdot]$. If $T_i$, $i=1,\dots, \dim \mathfrak g$, is a basis we write
\begin{equation}
[T_i, T_j ]= f_{\hph{k}ij}^{k} T_k,
\end{equation}
where $f_{\hph{k}ij}^{k}$ are the structure constants of $\mathfrak g$.
Given two elements $x,y \in \mathfrak g$ expanded in this basis,
$x = x^l T_l$, $y=y^m T_m$, their Lie bracket is $[x,y]=[x,y]^kT_k$ with
\begin{equation}
[x,y]^k = f_{\hph{k}lm}^k x^l y^m.
\end{equation}
Given then two elements
\begin{equation}
x^k_{a_1\cdots a_p}\in\Om^p\otimes\mathfrak{g}, \qquad
y^k_{a_1\cdots a_q}\in\Om^q\otimes\mathfrak{g}
\end{equation}
This can be extended to the map
\begin{equation}
[\cdot,\cdot]:\lp\Om^p\otimes\mathfrak{g}\rp\times\lp\Om^q\otimes\mathfrak{g}\rp\longrightarrow\Om^{p+q}\otimes\mathfrak{g},
\end{equation}
by using the wedge product on the first factors,
\begin{equation}\label{e16}
[x,y]^k_{a_1\cdots a_{p+q}}=\frac{\lp p+q\rp!}{p!q!}f^k_{\hph{k}l m}x^l_{[a_1\cdots a_p}y^m_{a_{p+1}\cdots a_{p+q}]}.
\end{equation}
The antisymmetry of the structure constants in eqn.\ (\ref{eq:fClosure}) amounts to
\begin{equation}
 [x,y]=-\lp -1\rp^{pq}[x,y],
\end{equation}
while the Jacobi identity becomes
\begin{equation}
 \lp -1\rp^{pr}[x,[y,z]]+\lp -1\rp^{pq}[y,[z,x]]+\lp -1\rp^{qr}[z,[x,y]]=0.
\end{equation}
Here $p$, $q$, and $r$ are the spacetime degrees of $x$, $y$, and $z$ respectively.

Then there are maps, denoted by $t$, which furnish a representation of $\mathfrak{g}$ on the complex $V_\bullet$.  In other words, given an element $x\in\mathfrak{g}$,
\begin{equation}
t_x:   V_p \to V_p,
\end{equation}
is a linear map which respects the Lie bracket. eqn.\ (\ref{e214a}) then becomes
\begin{equation}\label{e312}
t_x t_y-t_y t_x=t_{[x,y]}, \qquad \forall x , y \in \algg ,
\end{equation}
Using the notation $t_x(\varphi)=t(x,\varphi)$, then $t$ is also linear in its first argument.

The map $t$ can also be extended to
\begin{equation}
t:\lp\Om^p\otimes\mathfrak{g}\rp\times\lp\Om^q\otimes V_r\rp\longrightarrow\Om^{p+q}\otimes V_r,
\end{equation}
by acting with $t$ on the second factors as before, and with a wedge product on the first factors.  Explicitly, if
\begin{equation}
x^k_{a_1\cdots a_p}\in\Om^p\otimes\mathfrak{g},\qquad\varphi^{I_r}_{b_1\cdots b_q}\in\Om^q\otimes V_r,
\end{equation}
then
\begin{equation}
t(x,\varphi)^{I_r}_{a_1\cdots a_{p+q}}=\frac{\lp p+q\rp!}{p!q!}\lp t_k\rp^{I_r}_{\hph{I_r}J_r}x^k_{[a_1\cdots a_p}\varphi^{J_r}_{a_{p+1}\cdots a_{p+q}]}.
\end{equation}
The closure equation, (\ref{e214a}), becomes
\begin{equation}\label{ee317}
t_x t_y-\lp -1\rp^{pq}t_x t_ y-t_{[x,y]}=0,\qquad \forall x,y\in \algg, \varphi \in V_\bullet ,
\end{equation}
where $p$ and $q$ are the spacetime degrees of $x$ and $y$.

The next closure condition, eqn.\ (\ref{e214b}), now takes the form
\begin{equation}
\label{e317}
t_x q =q t_x,\qquad \forall x\in \algg, \varphi \in V_\bullet ,
\end{equation}
This is the statement that the diagram
\begin{displaymath}
\xymatrix{
\dots \ar[r] &
	V_{p+1} \ar[r]^{q} \ar[d]_{t_x}&
	V_p \ar[r]^{q} \ar[d]_{t_x}&
	V_{p-1} \ar[r]^{q} \ar[d]_{t_x} & \dots\\
\dots \ar[r] &
	V_{p+1} \ar[r]^{q} &
	V_p \ar[r]^{q} &
	V_{p-1} \ar[r]^{q} & \dots
}
\end{displaymath}
is commutative and $t_x: V_\bullet \to V_\bullet $ is a chain map.
Technically this says that the chain complex $V_\bullet$ with boundary operator $q$ is equivariant with respect to the action of $\mathfrak{g}$ encoded by $t$.

Finally, given a $x \in \algg$ we define the linear map
\begin{equation}
h_x:  V_{p-1} \to V_{p},
\end{equation}
which in the notation $h_x(\varphi) = h(x, \varphi)$, $\varphi \in V_{p-1}$,
are also linear in their first arguments.
Diagrammatically
\begin{displaymath}
\xymatrix{
\dots \ar[r] &
	V_{p+1} \ar[r] \ar@<+.5ex>[d]^{t_x} \ar[dl]_{h_x}&
	V_p \ar[r] \ar@<+.5ex>[d]^{t_x} \ar[dl]_{h_x}&
	V_{p-1} \ar[r]\ar@<+.5ex>[d]^{t_x} \ar[dl]_{h_x}& \ar[dl]_{h_x}\dots\\
\dots \ar[r] &
	V_{p+1} \ar[r] &
	V_p \ar[r] &
	V_{p-1} \ar[r] & \dots
}
\end{displaymath}
This can also be extended to a product
\begin{equation}
h:\lp\Om^p\otimes\mathfrak{g}\rp\times\lp\Om^q\otimes V_r\rp\longrightarrow\Om^{p+q}\otimes V_{r+1},
\end{equation}
via
\begin{equation}
h(x,\varphi)^{I_{r+1}}_{a_1\cdots a_{p+q}}=\frac{\lp p+q\rp!}{p!q!}\lp h_k\rp^{I_{r+1}}_{\hph{I_{r+1}}J_r}x^k_{[a_1\cdots a_p}\varphi^{J_r}_{a_{p+1}\cdots a_{p+q}]}.
\end{equation}

Condition (\ref{e217b}) then states
\begin{equation}
\label{e321}
q h_x+h_x q+t_x=0.
\end{equation}
This says that the linear map $t_x$ is chain-homotopic to the zero map.

We write the closure condition (\ref{e214c}) as two equations
\begin{subequations}\label{e323}
\begin{align}
\label{e323a}
& h_{[x,y]}= {1\o 2}\left[
h_x t_ y-\lp -1\rp^{pq}t_y h_x   +t_x h_y -(-1)^{pq} h_y t_x \right],\\
\label{e323b}
& t_x h_y +(-1)^{pq} t_y h_x= h_x t_y +(-1)^{pq} h_y t_x,
\end{align}
\end{subequations}
where $p$ and $q$ are the spacetime degrees of $x$ and $y$. In the first equation the symmetries of the Lie bracket and the Jacobi identity are manifest.

Gauge invariance (eqn.\ (\ref{e217c})), also requires
\begin{equation}
\label{e322}
h_ x h_ y +\lp -1\rp^{pq}h_y h_x=0.
\end{equation}

\subsection{Interior Product and Lie Derivative}\label{intprod}

In the case of dimensional reduction, we have an especially nice interpretation.  We will re\"examine this story in slightly more detail in section~\ref{sec:DimRed}, but the reader might find a preview of the discussion to be useful here.
For illustrative purposes consider the compactification from eleven to four dimensions on a seven-dimensional manifold $M$. In ref.\ \cite{Becker:2014uya} we found (adapted to the notation of the present paper)
\begin{equation} \label{dcsai}
\begin{split}
\d {\tt C}_{ijk}& = 3 \pa_{[i} {\widetilde \L}  _{jk]}+\xi^l\pa_l{\tt C}_{ijk}+3\pa_{[i}\xi^l{\tt C}_{jk]l}, \\
\d {\tt C}_{a  ij}& =  {{\cal D} }_{a } {\widetilde \L} _{ij}+ 2 \pa_{[i} {\widetilde \L}_{j]a }+\xi^k\pa_k{\tt C}_{aij}+2\pa_{[i}\xi^k{\tt C}_{|a|j]k},\\
\d {\tt C}_{a  b  i}& =
2 {{\cal D} }_{[a } {\widetilde \L}  _{b ]i }+\pa_i {\widetilde \L}_{a b }-{\widetilde \L}  _{ij}
{\cal F}_{a b }^j +\xi^j\pa_j{\tt C}_{abi}+\pa_i\xi^j{\tt C}_{abj} ,\\
\d {\tt C}_{a  b  c}& =3 {{\cal D} }_{[a } {\widetilde \L}  _{b  c ]}+
3 {\widetilde \L}  _{i[a } {\cal F}_{b c]}^i+\xi^i\pa_i{\tt C}_{abc} ,\\
\end{split}
\end{equation}
where the covariant derivative is defined by
\begin{equation}
\begin{split}
{{\cal D} }_{a_1 }{\widetilde \L}_{a_2 \dots a_n  i_1 \dots i_{p-n} }  =& 
\pa_{a_1 }{\widetilde \L}_{a_2 \dots a_n  i_1 \dots i_{p-n} }
-{\cal A}^a_{a_1} \pa_a {\widetilde \L}_{a_2 \dots a_n i_1 \dots i_{p-n}} \\
& + (p-n) (-1)^{p-n}{\widetilde \L}_{a_2 \dots a_n a [i_1 \dots i_{p-n-1}}\pa_{i_{p-n}]} {\cal A}_{a_1}^a.
\end{split}
\end{equation}

The Lie algebra $\mathfrak{g}$ is the algebra of tangent vector fields $x^k$ on the internal space $M$ (i.e.\ $\mathfrak{g}\cong\G(TM)$).  The bracket is the Lie bracket on tangent vector fields.  The chain complex is the (dual of) the de Rham complex on $M$, $V_p\cong\Om^{n-p}(M)$, and the operator $q$ is (up to a sign) the exterior derivative $d_M$ on $M$.  The representation $t$ is the Lie derivative, so $t(x,\varphi)$ becomes $\mathcal{L}_x\varphi$ for a tangent vector field $x$ and a differential form $\varphi\in\Om^\bullet(M)$.  Finally, the operator $h$ is contraction, so $h(x,\varphi)$ becomes $\iota_x\varphi$, again up to a sign.

Using this language, some of the equations in section \ref{abs} include some fairly famous equations.
So for example, eqn.\ (\ref{e312}) is
\begin{equation}
\mathcal{L}_x\mathcal{L}_y-\mathcal{L}_y\mathcal{L}_x = \mathcal{L}_{[x,y]}.
\end{equation}
while eqn.\ (\ref{e317}) is
\begin{equation}
d_M\mathcal{L}_x-\mathcal{L}_xd_M = 0
\end{equation}
while eqn.\ (\ref{e32}) is
\begin{equation}
d_M^2 = 0.
\end{equation}
Moreover, eqn.\ (\ref{e321}) is Cartan's magic formula
\begin{equation}
 \mathcal{L}_x=d_M\iota_x+\iota_xd_M .
\end{equation}
and eqn.\ (\ref{e322}) is the anti-symmetry of the interior product
\begin{equation}
 \iota_x\iota_y+\iota_y\iota_x = 0,
\end{equation}
while eqns.\ (\ref{e323a}) and (\ref{e323b}) correspond
\begin{equation}
{\cal L}_x\iota_y-\iota_y{\cal L}_x =
\iota_x\mathcal{L}_y-\mathcal{L}_y\iota_x =
\iota_{[x,y]}.
\end{equation}

\subsection{Covariant Derivatives and Bianchi Identities}

It is also useful to define a covariant exterior derivative
\begin{equation}
{{\cal D} }:\Om^p\otimes V_q\longrightarrow\Om^{p+1}\otimes V_q
\end{equation}
by ${{\cal D} }\varphi=d\varphi-t_{\cal A}\varphi$ or explicitly
\begin{equation}
\lp{{\cal D} }\varphi\rp^{I_q}_{a_1\cdots a_{p+1}}=\lp p+1\rp\left[ \pa_{[a_1}\varphi^{I_q}_{a_2\cdots a_{p+1}]}-\lp t_k\rp^{I_q}_{\hph{I_q}J_q}{\cal A}^k_{[a_1}\varphi^{J_q}_{a_2\cdots a_{p+1}]}\right].
\end{equation}

Then the variation and field strength for the matter fields becomes
\begin{equation}
\d\phi_{[p]}=t_{\l}\phi_{[p]}+{{\cal D} }\L _{[p-1]}+q(\L_{[p]})+
h_{{\cal F}}\L_{[p-2]},
\end{equation}
and
\begin{equation}
\begin{split}
& {\cal F}=d{\cal A}-\hlf\ls{\cal A},{\cal A}\rs,\\
& F_{[p+1]}={{\cal D} }\phi_{[p]}-q(\phi_{[p+1]})-h_{{\cal F}}\phi_{[p-1]},\\
& \d F_{[p+1]}=t_{\l } F_{[p+1]}.
\end{split}
\end{equation}
Here we have explicitly indicated spacetime degree with subscripts, and $\l=\l_{[0]}$, ${\cal A} =
{\cal A}_{[1]}$, ${\cal F} = {\cal F}_{[2]}$.

Next we can define the operator
\begin{equation}
{\cal Q} :\bigoplus_p\lp\Om^{p+q}\otimes V_p\rp\longrightarrow\bigoplus_p\lp\Om^{p+q+1}\otimes V_p\rp,
\end{equation}
for each $q$, via
\begin{equation}
{\cal Q}\varphi={{\cal D} }\varphi-\lp -1\rp^q\left[ q(\varphi)+h_{\cal F}\varphi\right],
\end{equation}
where $\varphi$ is an element of the direct sum above.  With this definition we have
\begin{equation}
{\cal Q}^2=0.
\end{equation}
In terms of this operator we have
\begin{equation}
\d\phi=t_\l \phi+{\cal Q}\L ,\qquad
F={\cal Q}\phi,\qquad
{\cal Q}F=0.
\end{equation}

\section{Supersymmetric Hierarchy}\label{susyhie}

In this section, we embed the gauged bosonic tensor hierarchy into 4D, $N=1$ superspace. The result is a gauged version of the abelian superspace hierarchy of ref.\ \cite{Becker:2016xgv}.

\subsection{Non-Abelian Gauge Symmetry in Superspace}\label{nonabsu} 

In this section we set up our conventions and derive some results which will be needed in forthcoming sections. As a result we keep some equations explicit. Section \ref{1213} parallels chapters 12 and 13 of ref.\ \cite{Wess:1992cp} and establishes some of our conventions. Unlike in the last section, we will write all spacetime vector and spinor indices explicitly, but promote all fields to superfields again valued in either $\mathfrak{g}$ or (in the next section) in $V_p$ for some $p$.  All the operations, $[\cdot,\cdot]$, $t$, $q$, and $h$ will be promoted to superfields in the obvious way, treating the fields as zero-forms (since we are writing the spacetime indices explicitly).  The one caveat is that anywhere that had a sign which depended on form degrees ({\it e.g.}\ a $(-1)^{pq}$), we will now have a sign in the case that both fields are anticommuting.

\subsubsection{$\algg$-Valued Superfields}\label{1213}

We first promote the gauge field $\mcA$ to a $\mathfrak{g}$-valued super-one-form $A_A$, {\it i.e.}\ a spinor-valued superfield of each chirality, $A_\al$ and its complex conjugate $\ov{A}_{\dot\al}$, and a real vector valued superfield $A_a$, all of which are also valued in $\mathfrak{g}$. We use capital
letters from the beginning of the Latin alphabet to label superspace coordinates.

Of course, there are far too many components included in these superfields.
Some of them can be removed by gauge transformations.  We would
like these to mimic the bosonic case, {\it i.e.}
\be
\d A_\al=D_\al\la+[\la,A_\al],\qquad\d\ov{A}_{\dot\al}=\ov{D}_{\dot\al}\la+[\la,\ov{A}_{\dot\al}],\qquad\d A_a=\pa_a\la+[\la,A_a],
\ee
for a real scalar superfield $\la$.  Here we are using the Lie bracket defined in eqn.\ (\ref{e16}).

In analogy with the bosonic case, we can build gauge-covariant combinations which assemble into a super-two-form $F_{AB}$,
\begin{subequations}
\begin{align}
F_{\al\beta} &= 2D_{(\al}A_{\beta)}-[A_\al,A_\beta],\\
F_{\dot\al\dot\beta} &= 2\ov{D}_{(\dot\al}\ov{A}_{\dot\beta)}-[\ov{A}_{\dot\al},\ov{A}_{\dot\beta}],\\
F_{\al\dot\beta} &= D_\al\ov{A}_{\dot\beta}+\ov{D}_{\dot\beta}A_\al-[A_\al,\ov{A}_{\dot\beta}]+2i\s^a_{\al\dot\beta}A_a,\\
F_{a\beta} &= \pa_aA_\beta-D_\beta A_a-[A_a,A_\beta],\\
F_{a\dot\beta} &= \pa_a\ov{A}_{\dot\beta}-\ov{D}_{\dot\beta}A_a-[A_a,\ov{A}_{\dot\beta}],\\
F_{ab} &= 2\pa_{[a}A_{b]}-[A_a,A_b].
\end{align}
\end{subequations}
These are covariant in the sense that
\be
\d F_{AB}=[\la,F_{AB}].
\ee

It is useful to define covariant derivatives which act on $\mathfrak{g}$-valued superfields,
\be
\mcD_\al x=D_\al x-[A_\al,x],\qquad\ov{\mcD}_{\dot\al}x=\ov{D}_{\dot\al}x-[\ov{A}_{\dot\al},x],\qquad\mcD_ax=\pa_ax-[A_a,x].
\ee
From these definitions it follows
\be
\label{eq:AVar}
\d A_\al=\mcD_\al\la,\qquad\d\ov{A}_{\dot\al}=\ov{\mcD}_{\dot\al}\la,\qquad\d A_a=\mcD_a\la.
\ee
In the next section we will also extend the action of the covariant derivative
to superfields from the tensor hierarchy.

By construction, the field strengths satisfy a number of Bianchi identities,
\begin{subequations}
\begin{align}
0 &= 3\mcD_{(\al}F_{\beta\g)},\\
0 &= 3\ov{\mcD}_{(\dot\al}F_{\dot\beta\dot\g)},\\
0 &= 2\mcD_{(\al}F_{\beta)\dot\g}+\ov{\mcD}_{\dot\g}F_{\al\beta}+4i\s^a_{(\al|\dot\g}F_{a|\beta)},\\
0 &= 2\ov{\mcD}_{(\dot\al}F_{|\g|\dot\beta)}+\mcD_\g F_{\dot\al\dot\beta}+4i\s^a_{\g(\dot\al}F_{|a|\dot\beta)},\\
0 &= \mcD_aF_{\al\beta}-2\mcD_{(\al}F_{|a|\beta)},\\
0 &= \mcD_aF_{\dot\al\dot\beta}-2\ov{\mcD}_{(\dot\al}F_{|a|\dot\beta)},\\
0 &= \mcD_aF_{\al\dot\beta}-\mcD_\al F_{a\dot\beta}-\ov{\mcD}_{\dot\beta}F_{a\al}-2i\s^b_{\al\dot\beta}F_{ab},\\
0 &= 2\mcD_{[a}F_{b]\al}+\mcD_\al F_{ab},\\
0 &= 2\mcD_{[a}F_{b]\dot\al}+\ov{\mcD}_{\dot\al} F_{ab},\\
0 &= 3\mcD_{[a}F_{bc]}.
\end{align}
\end{subequations}
Note that we are not imposing these identities; they are simply consequences of the definitions of the $F_{AB}$.

Even with the gauge transformations, however, there are still too many components in $A_A$.  Some extra conditions have to be imposed on the superfields. These conditions should be gauge
covariant, and should consequently be formulated in terms of $F_{AB}$.  We start by setting
\begin{equation}
F_{\al\beta}=F_{\dot\al\dot\beta}=F_{\al\dot\beta}=0.
\end{equation}
The last equation can be used to solve for $A_a$,
\be
A_a=-\frac{i}{4}\lp\ov{\s}_a\rp^{\dot\al\al}\lp D_\al\ov{A}_{\dot\al}+\ov{D}_{\dot\al}A_\al-[A_\al,\ov{A}_{\dot\al}]\rp.
\ee
Since $F_{a\al}$ splits into two irreducible representations of the four-dimensional Lorentz group, of spin $1/2$ and spin $3/2$, we next set the spin $3/2$ piece to zero.  Explicitly, this means
\be
\s^a_{(\al|\dot\al}F_{a|\beta)}=0,
\ee
and its complex conjugate,
\be
\s^a_{\al(\dot\al}F_{|a|\dot\beta)}=0.
\ee
The remaining components of $F_{a\al}$ are captured by
\begin{equation}
\begin{split}
\mcW^\al & =-\frac{1}{4}\lp\ov{\s}^a\rp^{\dot\al\al}F_{a\dot\al},\\
\ov{\mcW}^{\dot\al}& =-\frac{1}{4}\lp\ov{\s}^a\rp^{\dot\al\al}F_{a\al},
\end{split}
\ee
or equivalently
\be
\begin{split}
F_{a\al}& =\lp\s_a\rp_{\al\dot\al}\ov{\mcW}^{\dot\al},\\ F_{a\dot\al}& =\lp\s_a\rp_{\al\dot\al}\mcW^\al.
\end{split}
\ee
$F_{ab}$ is determined by the Bianchi identity to be
\be
F_{ab}=-\frac{i}{2}\left[ \lp\s_{ab}\rp_\al^{\hph{\al}\beta}\mcD^\al\mcW_\beta-
\lp\ov{\s}_{ab}\rp^{\dot\al}_{\hph{\dot\al}\dot\beta}\ov{\mcD}_{\dot\al}\ov{\mcW}^{\dot\beta}
\right].
\ee
Here the normalization of $\mcW$ has been chosen to agree with the conventions of ref.\ \cite{Becker:2016xgv}. Taking the symmetric part
\be
\mcD^\al\mcW_\al-\ov{\mcD}_{\dot\al}\ov{\mcW}^{\dot\al}=0.
\ee
Finally, from a different Bianchi identity we have
\be
0=\ov{\mcD}_{(\dot\al}F_{|a|\dot\beta)}=\lp\s_a\rp_{\beta(\dot\beta}\ov{\mcD}_{\dot\al)}\mcW^\beta,
\ee
and contracting with $(\ov{\s}^a)^{\dot\beta\al}$, we learn that
\be
\ov{\mcD}_{\dot\al}\mcW^\al=0.
\ee
Of course, we can also derive the conjugate,
\be
\mcD_\al\ov{\mcW}^{\dot\al}=0.
\ee

Finally, we note that the covariant derivatives obey an algebra
\begin{subequations}
\begin{align}
2\mcD_{(\al}\mcD_{\beta)}x &= 0,\\
2\ov{\mcD}_{(\dot\al}\ov{\mcD}_{\dot\beta)}x &= 0,\\
\mcD_\al\ov{\mcD}_{\dot\al}x+\ov{\mcD}_{\dot\al}\mcD_\al x &= -2i\s^a_{\al\dot\al}\mcD_ax,\\
\mcD_a\mcD_\al x-\mcD_\al\mcD_ax &= -\lp\s_a\rp_{\al\dot\al}[\ov{\mcW}^{\dot\al},x],\\
\mcD_a\ov{\mcD}_{\dot\al}x-\ov{\mcD}_{\dot\al}\mcD_ax &= -\lp\s_a\rp_{\al\dot\al}[\mcW^\al,x],\\
2\mcD_{[a}\mcD_{b]}x &= -[F_{ab},x].
\end{align}
\end{subequations}
Some additional identities include (we define $\mcD^2=\mcD^\al\mcD_\al$ and $\ov{\mcD}^2=\ov{\mcD}_{\dot\al}\ov{\mcD}^{\dot\al}$, and we will always write $\mcD^a\mcD_a$ out explicitly to distinguish it from the $\mcD^2$ just defined)
\bea
\mcD^2\ov{\mcD}_{\dot\al}x-\ov{\mcD}_{\dot\al}\mcD^2x &=& -2i\s^a_{\al\dot\al}\lp\mcD_a\mcD^\al x+\mcD^\al\mcD_ax\rp,\\
\ov{\mcD}^2\mcD_\al x-\mcD_\al\ov{\mcD}^2x &=& 2i\s^a_{\al\dot\al}\lp\mcD_a\ov{\mcD}^{\dot\al}x+\ov{\mcD}^{\dot\al}\mcD_ax\rp,\\
\mcD^\al\ov{\mcD}^2\mcD_\al x-\ov{\mcD}_{\dot\al}\mcD^2\ov{\mcD}^{\dot\al}x &=& 8i\Om_{\mathfrak{g}}(\mcW,x),
\eea
where we have defined
\be
\Om_{\mathfrak{g}}(\psi,x)=[\psi^\al,\mcD_\al x]+[\ov{\psi}_{\dot\al},\ov{\mcD}^{\dot\al}x]+\hlf[\mcD^\al\psi_\al+\ov{\mcD}_{\dot\al}\ov{\psi}^{\dot\al},x],
\ee
as an operator on any $\mathfrak{g}$-valued covariantly chiral spinor superfield $\psi$ and a real $\mathfrak{g}$-valued superfield $x$.

\subsubsection{$V_p$-Valued Superfields}\label{vv1}

Now we will combine the hierarchy structure from the first sections with the non-abelian gauge superfield in the last section.  For a $V_p$-valued superfield $\varphi$, we define covariant derivatives
\be
\label{eq:tCovariantD}
\mcD_\al\varphi=D_\al\varphi-t(A_\al,\varphi),\qquad\ov{\mcD}_{\dot\al}\varphi=
\ov{D}_{\dot\al}\varphi-t(\ov{A}_{\dot\al},\varphi),\qquad\mcD_a\varphi=
\pa_a\varphi-t(A_a,\varphi).
\ee
These satisfy an algebra
\begin{subequations}
\begin{align}
2\mcD_{(\al}\mcD_{\beta)}\varphi &= 0,\\
2\ov{\mcD}_{(\dot\al}\ov{\mcD}_{\dot\beta)}\varphi &= 0,\\
\mcD_\al\ov{\mcD}_{\dot\al}\varphi+\ov{\mcD}_{\dot\al}\mcD_\al\varphi &= -2i\s^a_{\al\dot\al}\mcD_a\varphi,\\
\mcD_a\mcD_\al\varphi-\mcD_\al\mcD_a\varphi &= -\lp\s_a\rp_{\al\dot\al}t(\ov{\mcW}^{\dot\al},\varphi),\\
\mcD_a\ov{\mcD}_{\dot\al}\varphi-\ov{\mcD}_{\dot\al}\mcD_a\varphi &= -\lp\s_a\rp_{\al\dot\al}t(\mcW^\al,\varphi),\\
2\mcD_{[a}\mcD_{b]}\varphi &= -t(F_{ab},\varphi).
\end{align}
\end{subequations}
Also,
\begin{subequations}
\begin{align}
\mcD^2\ov{\mcD}_{\dot\al}\varphi-\ov{\mcD}_{\dot\al}\mcD^2\varphi &= -2i\s^a_{\al\dot\al}\lp\mcD_a\mcD^\al\varphi+\mcD^\al\mcD_a\varphi\rp,\\
\ov{\mcD}^2\mcD_\al\varphi-\mcD_\al\ov{\mcD}^2\varphi &= 2i\s^a_{\al\dot\al}\lp\mcD_a\ov{\mcD}^{\dot\al}\varphi+\ov{\mcD}^{\dot\al}\mcD_a\varphi\rp,\\
\label{e425c}
\mcD^\al\ov{\mcD}^2\mcD_\al\varphi-\ov{\mcD}_{\dot\al}\mcD^2\ov{\mcD}^{\dot\al}\varphi &= 8i\Om_t(\mcW,\varphi),
\end{align}
\end{subequations}
where we defined
\be
\Om_t(\mcW,\varphi)=t(\mcW^\al,\mcD_\al\varphi)+t(\ov{\mcW}_{\dot\al},\ov{\mcD}^{\dot\al}\varphi)+\hlf t(\mcD^\al\mcW_\al+\ov{\mcD}_{\dot\al}\ov{\mcW}^{\dot\al},\varphi).
\ee
Note that the last term can be rewritten using $\mcD^\al\mcW_\al=\ov{\mcD}_{\dot\al}\ov{\mcW}^{\dot\al}$.

These covariant derivatives also have many nice properties with respect to the operators $t$, $q$, and $h$.  In particular,
\begin{subequations}
\begin{align}
\mcD_\al t(x,\varphi)& =t(\mcD_\al x,\varphi)\pm t(x,\mcD_\al\varphi),\\
\quad\ov{\mcD}_{\dot\al}t(x,\varphi)& =t(\ov{\mcD}_{\dot\al}x,\varphi)\pm t(x,\ov{\mcD}_{\dot\al}\varphi),\\\
\mcD_at(x,\varphi)& =t(\mcD_ax,\varphi)+t(x,\mcD_a\varphi),\\
\mcD_\al q(\varphi)& =q(\mcD_\al\varphi),\quad\ov{\mcD}_{\dot\al}q(\varphi)=q(\ov{\mcD}_{\dot\al}\varphi),\\
\mcD_aq(\varphi)& =q(\mcD_a\varphi),\\
\mcD_\al h(x,\varphi)& =h(\mcD_\al x,\varphi)\pm h(x,\mcD_\al\varphi),\\
\ov{\mcD}_{\dot\al}h(x,\varphi)& =h(\ov{\mcD}_{\dot\al}x,\varphi)\pm h(x,\ov{\mcD}_{\dot\al}\varphi),\\
\mcD_ah(x,\varphi)& =h(\mcD_ax,\varphi)+h(x,\mcD_a\varphi),
\end{align}
\end{subequations}
where the upper sign is for $x$ being a commuting superfield, and the lower sign is for $x$ being anticommuting.

\subsubsection{Chern-Simons Superfield}\label{vv2}

In this subsection we define an operator $\Om_h$ which takes a $\mathfrak{g}$-valued covariantly chiral spinor superfield $\psi$ (in practice $\psi$ will always be $\mcW$) and a $V_p$-valued scalar superfield $\varphi$, and returns a $V_{p+1}$-valued scalar superfield,
\be
\Om_h(\psi,\varphi)=h(\psi^\al,\mcD_\al\varphi)+h(\ov{\psi}_{\dot\al},\ov{\mcD}^{\dot\al}\varphi)+\hlf h(\mcD^\al\psi_\al+\ov{\mcD}_{\dot\al}\ov{\psi}^{\dot\al},\varphi).
\ee
This satisfies
\bea
-\frac{1}{4}\ov{\mcD}^2\Om_h(\psi,\varphi) &=& h(\psi^\al,-\frac{1}{4}\ov{\mcD}^2\mcD_\al\varphi)-\frac{1}{8}\ov{\mcD}^2h(\mcD^\al\psi_\al-\ov{\mcD}_{\dot\al}\ov{\psi}^{\dot\al},\varphi),\\
-\frac{1}{4}\mcD^2\Om_h(\psi,\varphi) &=& h(\ov{\psi}_{\dot\al},-\frac{1}{4}\mcD^2\ov{\mcD}^{\dot\al}\varphi)+\frac{1}{8}\mcD^2h(\mcD^\al\psi_\al-\ov{\mcD}_{\dot\al}\ov{\psi}^{\dot\al},\varphi).
\eea
For the case of $\psi=\mcW$ the second terms drop out and we have
\be
-\frac{1}{4}\ov{\mcD}^2\Om_h(\mcW,\varphi)=h(\mcW^\al,-\frac{1}{4}\ov{\mcD}^2\mcD_\al\varphi),\quad -\frac{1}{4}\mcD^2\Om_h(\mcW,\varphi)=h(\ov{\mcW}_{\dot\al},-\frac{1}{4}\mcD^2\ov{\mcD}^{\dot\al}\varphi).
\ee
Note also that $\Om_h$ inherits certain properties from $h$, for instance from (\ref{e214c}),
\be\label{e432}
\Om_h(\psi,t(x,\varphi))-t(x,\Om_h(\psi,\varphi))+\Om_h([x,\psi],\varphi)=0,
\ee
and from (\ref{e217b}),
\be\label{e433}
q(\Om_h(x,\varphi))+\Om_h(x,q(\varphi))+\Om_t(x,\varphi)=0.
\ee
Finally, note that if $\varphi$ is covariantly chiral, then
\be
\label{e434}
\Om_h(\psi,\varphi)=h(\mcW^\al,\mcD_\al\varphi)+h(\mcD^\al\mcW_\al,\varphi),
\ee
while if $\varphi$ is antichiral,
\be\label{e435}
\Om_h(\psi,\varphi)=h(\ov{\mcW}_{\dot\al},\ov{\mcD}^{\dot\al}\varphi)+h(\ov{\mcD}_{\dot\al}\ov{\mcW}^{\dot\al},\varphi).
\ee

\subsection{Incorporating the Hierarchy}\label{inchie}

\subsubsection{Prepotentials}\label{prepot} 
The hierarchy consists of the following components and their prepotential superfields \cite{Gates:1980ay,Becker:2016xgv}:
\begin{itemize}
\item A collection of $V_0$-valued covariantly chiral superfields $\Phi$, i.e.\
\be
\ov{\mcD}_{\dot\al}\Phi=0.
\ee
The axions are given by
\be
a=\hlf\lp\Phi+\ov{\Phi}\rp\Big|.
\ee
The vertical slash means that we evaluate the superfield at $\th=\bar{\th}=0$, i.e.\ we take the lowest component.
\item A collection of $V_1$-valued real superfields $V$.  We have
\be
A_a=-\frac{1}{4}\ov{\s}_a^{\dot\al\al}\lp\mcD_\al\ov{\mcD}_{\dot\al}-\ov{\mcD}_{\dot\al}\mcD_\al\rp V\Big|.
\ee
Note that this map to components now depends on the non-abelian gauge field!
\item A collection of $V_2$-valued covariantly chiral spinor superfields $\Sigma_\al$, $\ov{\mcD}_{\dot\al}\Sigma_\al=0$.  We also have
\be
B_{ab}=-\frac{i}{2}\lp\lp\s_{ab}\rp_\al^{\hph{\al}\beta}\mcD^\al\Sigma_\beta-\lp\ov{\s}_{ab}\rp^{\dot\al}_{\hph{\dot\al}\dot\beta}\ov{\mcD}_{\dot\al}\ov{\Sigma}^{\dot\beta}\rp\Big|.
\ee
\item A collection of $V_3$-valued real superfields $X$.
\be
C_{abc}=\frac{1}{8}\e_{abcd}\lp\ov{\s}^d\rp^{\dot\al\al}\lp\mcD_\al\ov{\mcD}_{\dot\al}-\ov{\mcD}_{\dot\al}\mcD_\al\rp X\Big|.
\ee
\item A collection of $V_4$-valued covariantly chiral superfields $\G$, $\ov{\mcD}_{\dot\al}\G=0$.
\be
D_{abcd}=\frac{i}{8}\lp\mcD^2\G-\ov{\mcD}^2\ov{\G}\rp\Big|.
\ee
\end{itemize}

Now we declare the following variations
\begin{subequations}
\label{eqs:SuperfieldVariations}
\begin{align}
\label{eq:PhiVar}
\d\Phi &= t_\la \Phi+q(\La),\\
\d V &= t_\la V+\frac{\La-\ov{\La}}{2i}+q(U),\\
\d\Sigma_\al &= t_\la \Sigma_\al-\frac{1}{4}\ov{\mcD}^2\mcD_\al U+q(\Upsilon_\al)+h_{\mcW_\al}\La,\\
\d X &= t_\la X+{1\o 2i}\left( {\mcD^\al\Upsilon_\al-\ov{\mcD}_{\dot\al}\ov{\Upsilon}^{\dot\al}}\right)
+q(\Xi)+\Om_h(\mcW,U),\\
\d\G &= t_\la \G-\frac{1}{4}\ov{\mcD}^2\Xi+h_{\mcW^\al}\Upsilon_\al.
\end{align}
\end{subequations}
Here in addition to the $\mathfrak{g}$-valued real superfield $\la$, we have gauge parameters $\La$, which is a $V_1$-valued covariantly chiral superfield, $U$, which is a $V_2$-valued real scalar superfield, $\Upsilon_\al$, a $V_3$-valued covariantly chiral spinor superfield, and $\Xi$, a $V_4$-valued real scalar superfield.

Covariantly chiral fields remain so after a gauge transformation, i.e.\ given
covariantly chiral fields $\Phi$, $\Sigma_\a$ and $\Gamma$
\begin{equation}
\d\lp\ov{\mcD}_{\dot\al}\Phi\rp=0, \qquad
\d\lp\ov{\mcD}_{\dot\al}\Sigma_\al\rp=0,\qquad
\d\lp\ov{\mcD}_{\dot\al}\G\rp=0.
\end{equation}

Finally, note also that we can go to a Wess-Zumino-like gauge for each of these transformations.  After this gauge fixing, the only residual gauge symmetries are the bosonic ones with parameters $\La^{I_p}_{a_1\cdots a_{p-1}}$.  In this gauge, the transformations of the components $a^A$, $A^I_a$, $B^M_{ab}$, $C^S_{abc}$ and $D^X_{abcd}$, defined in this section, simply match eqn.\ (\ref{e27}).

\subsubsection{Field Strengths}\label{fieldstr}

Next we define field strengths
\begin{subequations}
\label{eqs:FieldStrengths}
\begin{align}
E &= -q(\Phi),\\
F &= {1\o 2i}\left( {\Phi-\ov{\Phi}}\right)-q(V),\\
W_\al &= -\frac{1}{4}\ov{\mcD}^2\mcD_\al V-q(\Sigma_\al)-h_{\mcW_\al}\Phi,\\
H &={1\o 2i} \left( {\mcD^\al\Sigma_\al-\ov{\mcD}_{\dot\al}\ov{\Sigma}^{\dot\al}}\right) -q(X)-\Om_h(\mcW,V),\\
G &= -\frac{1}{4}\ov{\mcD}^2X-q(\G)-h_{\mcW^\al}\Sigma_\a.
\end{align}
\end{subequations}

We can check that these are covariant, making heavy use of eqns.\ (\ref{ee317}), (\ref{e317}), (\ref{e321}), (\ref{e323}), (\ref{e322}), as well as the algebra of the $\mcD$'s, and the way they commute through the operators $t$, $q$, and $h$. We denote these as the hierarchy equations. So, for example, very explicitly
\begin{equation}
\d E= -q\left[ t_\la \Phi+q(\La)\right]
\stackrel{(\ref{e32})}{=}
-q t_\l (\Phi)
\stackrel{(\ref{e317})}{=} t_\la E.
\end{equation}
or
\begin{equation}
\d F  =    {1\o 2 i }\left(t_\l \Phi-t_\l \ov{\Phi}\right)
-q\left[ t_\l V+q(U)\right]
 \stackrel{(\ref{e32})}{=} {1\o 2 i }\left(t_\l \Phi-t_\l \ov{\Phi}\right)
-q t_\l V
\stackrel{(\ref{e317})}{=}  t_\l F.
\end{equation}
Here we have dropped terms which cancel trivially. However, terms which only cancel after using the hierarchy equations are kept explicit and the equation being used is
indicated.  In the same way

\begin{equation}
\begin{split}
\d W_\al = &  t_\l \left[ - \frac{1}{4}\ov{\mcD}^2\mcD_\al V-q(\Sigma_\al)\right]-
\left(  t_{\mcW_\al}+q h_{\mcW_\al}+h_{\mcW_\al} q \right)\L\\
& -
\left(  h_{\mcW_\al} t_\l +h_{[\la,\mcW_\al]}\right)\Phi+ q^2 (\Upsilon _\a)= t_\l W_\al,
\end{split}
\end{equation}
Here eqns.\ (\ref{e32}), (\ref{e321}), (\ref{e323}) have been used. Moreover,
we have used the identity
\be
\frac{i}{8}\ov{\mcD}^2\mcD_\al\La=
-\frac{1}{4}\s^a_{\al\dot\al}\ov{\mcD}^{\dot\al}\mcD_a\La=
\frac{1}{4}\s^a_{\al\dot\al}\ov{\s}_a^{\dot\al\beta}t_{\mcW_\beta}\La=-t_{\mcW_\al}\La,
\ee
valid for any covariantly chiral field $\L$.

Next consider $H$. We find

\begin{equation}
\begin{split}
& \d H
= \frac{1}{2i}\lp t_\la (\mcD^\al\Sigma_\al)-
t_\la (\ov{\mcD}_{\dot\al}\ov{\Sigma}^{\dot\al}) \rp -q t_\la X -\Om_h([\l,{\cal W}],V) - \Om_h({\cal W}, t_\l V)\\
& +{1\o 2 i }\left( -\frac{1}{4}\mcD^\al\ov{\mcD}^2\mcD_\al U +
\frac{1}{4}\ov{\mcD}_{\dot\al}\mcD^2\ov{\mcD}^{\dot\al}U\right) -
\Om_h(\mcW,q(U))- q \Om _h({\cal W},U)-
q^ 2 \Xi\\
& + {1\o 2 i }\left(h(\mcD^\al\mcW_\al,\La)+h(\mcW^\al,\mcD_\al\La)-
h(\ov{\mcD}_{\dot\al}\ov{\mcW}^{\dot\al},\ov{\La})-
h(\ov{\mcW}_{\dot\al},\ov{\mcD}^{\dot\al}\ov{\La}) \right)  -\Om_h(\mcW,\frac{\La-\ov{\La}}{2i})\\
& = t_ \la H.
\end{split}
\end{equation}
Note that the second and third lines of this equation vanish after using eqns.\ (\ref{e32}), (\ref{e425c}), (\ref{e433}), (\ref{e434}), and (\ref{e435}). The first line
can be rewritten using eqns.\ \C{e317} and \C{e432}.

And finally consider
\begin{equation}
\begin{split}
\d G=& -\frac{1}{4}
t_\la (\ov{\mcD}^2X)-qt_\la \G-h_{[\la,\mcW^\al]} \Sigma_\a
- h_{\mcW^\al}t_\la \Sigma_\al\\
& +\frac{i}{8}\ov{\mcD}^2\mcD^\al\Upsilon_\al -q h_{\mcW^\al}\Upsilon_\al
- h_{\mcW^\al} q(\Upsilon_\al) -h_{\mcW^\al} h_{\mcW_\al}\La
\\
=& t_\la G.
\end{split}
\end{equation}
Here eqns.\ (\ref{e323}), (\ref{e317}) have been used.
Note that the last term $h_{\mcW^\al} h_{\mcW_\al}\La$, vanishes after taking into account
that the $\mcW$'s are anticommuting and we are contracting their indices with $\e_{\al\beta}$.  The combination of these two antisymmetries makes the result symmetric so we can use (\ref{e217c}).

To summarize the after a gauge transformation the superfield strengths change according to
\begin{equation}
\begin{split}
& \d E = t_\l E, \\
& \d F = t_\l F, \\
& \d W_\a = t_\l W_\a, \\
& \d H = t_\l H, \\
& \d G = t_\l G.  \\
\end{split}
\end{equation}

With similar manipulations we can show that the field strengths obey Bianchi identities,
\begin{subequations}
\begin{align}
0 &= q(E),\\
0 &= \frac{1}{2i} \left( E-\ov{E}\right) +q(F),\\
0 &= -\frac{1}{4}\ov{\mcD}^2\mcD_\al F+q(W_\al)+h_{\mcW_\al} E,\\
0 &= \frac{1}{2i}\left( \mcD^\al W_\al-\ov{\mcD}_{\dot\al}\ov{W}^{\dot\al}\right)+q(H)+\Om_h(\mcW,F),\\
0 &= -\frac{1}{4}\ov{\mcD}^2H+q(G)+h_{\mcW^\al} W_\al.
\end{align}
\end{subequations}

\section{Bosonic Chern-Simons Actions}
\label{sec:BCSAs}

Next we turn to the task of constructing gauge-invariant actions.  One possibility is simply to build a spacetime scalar $f(\mcF,F)$ out of our covariant field strengths $\mcF_{[2]}$ and $F_{[p+1]}$, and then take an action
\be
S=\int d^dxf(\mcF,F).
\ee
The condition for gauge invariance is simply that $f$ is also a singlet under the non-abelian gauge transformations, i.e.\ that schematically
\be
0=\d_\la f(\mcF,F)=f([\la,\mcF],F)+f(\mcF,t_\la F).
\ee
where it is understood that the last term should be expanded with one term for each $F^{I_p}_{[p+1]}$.

This is not the only way to construct a gauge-invariant action, however.  Another option is to have a Chern-Simons type action, in which the Lagrangian is not invariant, but rather transforms into a total derivative (so the action itself is invariant).  In this section we explore this possibility.

\subsection{Cohomological Interpretation of Abelian Bosonic Chern-Simons Actions}

We will proceed in the same way that we did in ref.~\cite{Becker:2016xgv}.  For us, a Chern-Simons action will be a sum of terms, each of which is the integral over spacetime of the wedge product of one potential and some number of field strengths.  For example, we can have a linear Chern-Simons term
\be
\label{eq:CS0Indices}
S_{0,CS}=\int\al_{I_d}\phi^{I_d}_{[d]},
\ee
or a quadratic Chern-Simons action
\be
\label{eq:CS1Indices}
S_{1,CS}=\int\sum_{p=0}^d\al_{I_pJ_{d-p-1}}\phi^{I_p}_{[p]}\w F^{J_{d-p-1}}_{[d-p]},
\ee
or a cubic Chern-Simons action
\be
\label{eq:CS2Indices}
S_{2,CS}=\int\sum_{p=0}^d\sum_{q=-1}^{\lfloor\frac{d-p}{2}\rfloor-1}\al_{I_pJ_qK_{d-p-q-2}}\phi^{I_p}_{[p]}\w F^{J_q}_{[q+1]}\w F^{K_{d-p-q-2}}_{[d-p-q-1]}.
\ee
In each case the coefficients $\al$ are just numbers which will have to satisfy certain identities in order for the action to be gauge-invariant.

In order to generalize this construction, we will introduce some notation.  For an element $\varphi\in V_p$, we can define its degree by
\be
\deg(\varphi)=p.
\ee
For a general element $\varphi\in V_\bullet$ there is not a well-defined degree unless we first project onto $V_p\subseteq V_\bullet$ with $\pi_p:V_\bullet\longrightarrow V_p$ (so $\deg(\pi_p(\varphi))=p$).

Now for fixed $N$, define a co-chain complex $X_{(N)}^\bullet$ via
\be
X_{(N)}^p=\bigoplus_{i_1+\cdots+i_N=p}\lp V_{i_1}^\ast\otimes\cdots\otimes V_{i_N}^\ast\rp,
\ee
where
\be
V_i^\ast=\Hom(V_i,\R)
\ee
is the dual space, and the co-boundary operator is defined by
\be
q:X_{(N)}^p\longrightarrow X_{(N)}^{p+1},
\ee
\be
\label{eq:qXDef}
\lp q\al\rp(\varphi_1,\cdots,\varphi_N)=\sum_{i=1}^N\lp -1\rp^{\sum_{j=1}^{i-1}\deg(\varphi_j)+i+1}\al(\varphi_1,\cdots,q(\varphi_i),\cdots,\varphi_N).
\ee
For example, for $\al\in X_{(3)}^p$, $\varphi_1\in V_i$, $\varphi_2\in V_j$, $\varphi_3\in V_{p-i-j+1}$, 
\begin{multline}
\lp q\al\rp(\varphi_1,\varphi_2,\varphi_3)\\
=\al(q(\varphi_1),\varphi_2,\varphi_3)+\lp -1\rp^{i+1}\al(\varphi_1,q(\varphi_2),\varphi_3)+\lp -1\rp^{i+j}\al(\varphi_1,\varphi_2,q(\varphi_3)).
\end{multline}
It is straightforward to check that $q^2=0$ on $X^\bullet_{(N)}$, using the fact that $q^2=0$ on $V_\bullet$.  


It will be useful to introduce some short hand notation.  For $\varphi_j\in V_\bullet$, which do not necessarily have well-defeined degrees, and for $\al\in X_{(N)}^p$, we will write
\be
\al(\varphi_1,\cdots,\varphi_N)\equiv\sum_{i_1+\cdots+i_N=p}\al(\pi_{i_1}(\varphi_1),\cdots,\pi_{i_N}(\varphi_N))\in\R.
\ee

We would also like to define a closely related sub-complex, $\widehat{X}_{(N)}^\bullet$, given by
\begin{multline}
\widehat{X}_{(N)}^p=\left\{\al\in X_{(N)}^p|\forall i_1,i_2,\mathrm{\ and\ }\varphi_1\in V_{i_1},\varphi_2\in V_{i_2},\varphi_0\in V_\bullet,\right.\non\\
\left.\mathrm{\ then\ }\al(\varphi_0,\cdots,\varphi_1,\cdots,\varphi_2,\cdots)=\lp -1\rp^{(i_1+1)(i_2+1)}\al(\varphi_0,\cdots,\varphi_2,\cdots,\varphi_1,\cdots)\right\},
\end{multline}
i.e.\ $\widehat{X}_{(N)}^\bullet$ consists of those $\al$ that are symmetric (with appropriate signs) in their final $N-1$ arguments.  We also have 
that for $\al\in\widehat{X}_{(N)}^p$, then $q\al\in\widehat{X}_{(N)}^{p+1}\subseteq X_{(N)}^{p+1}$, so $q$ is well-defined on $\widehat{X}^\bullet_{(N)}$.  Equivalently, there are obvious inclusion and projection maps between $X_{(N)}^\bullet$ and $\widehat{X}_{(N)}^\bullet$, and the various squares which combine these with $q$ commute.  Finally, note that this symmetrization is only non-trivial for $N>2$; in particular, $\widehat{X}_{(1)}^\bullet=X_{(1)}^\bullet$ and $\widehat{X}_{(2)}^\bullet=X_{(2)}^\bullet$.

We can easily extend the operation of any $\al$ to objects in $\Om^\bullet\otimes V_\bullet$ by combining it with the wedge product on differential forms, taken in the order of its arguments.  Fix an $N$, and pick some $\al\in \widehat{X}_{(N)}^{d+1-N}$.  Note that $\al(\phi,F,\cdots,F)$ (one potential, $N-1$ field strengths) is a $d$-form, so we can construct an action by integrating it over the space-time $\R^d$,
\be
S_{CS,\al}=\int\al(\phi,F,\cdots,F).
\ee
All of the previous examples of Chern-Simons terms (\ref{eq:CS0Indices}), (\ref{eq:CS1Indices}), and (\ref{eq:CS2Indices}), were of this form.  Note that this is the reason we defined the complex
$\widehat{X}_{(N)}^\bullet$.  If we took an $\al\in X_{(N)}^\bullet$ which was in the kernel of the projection onto $\widehat{X}_{(N)}^\bullet$ (the projection is simply symmetrization over the last $N-1$ arguments, so for example when $N=3$ the kernel consists of elements in $X_{(3)}^\bullet$ which are antisymmetric under exchange of the last two arguments), then the corresponding action would be zero for trivial reasons of symmetry.

Restricting for the moment to the abelian case only, what is the condition for gauge invariance?  The variation comes only from $\d\phi=d\La+q(\La)$.  After integrating by parts, using the Bianchi identities $dF=-q(F)$, and using the definition (\ref{eq:qXDef}), we have
\be
\d S_{CS,\al}=\int\al(d\La+q(\La),F,\cdots,F)=\int\lp q\al\rp(\La,F,\cdots,F).
\ee
We see immediately that a sufficient condition for gauge-invariance is that $\al$ is a closed element, $q\al=0$, of the co-chain complex $\widehat{X}^\bullet$, i.e.\ $\al$ is a cocycle.  

The necessary condition is actually a little bit weaker, since $\La$ and $F$ are not completely unconstrained elements of $V_\bullet$.  More explicitly, we can define projectors
\be
\pi_\La=\sum_{p=1}^d\pi_p,\quad\pi_\phi=\sum_{p=0}^d\pi_p,\quad\pi_F=\pi_{\operatorname{Im}(q)\cap V_{-1}}+\sum_{p=0}^{d-1}\pi_p,
\ee
where the first term in $\pi_F$ is the projection onto the image of the map $q:V_0\rightarrow V_{-1}$.  These projections simply capture the fact that $\La^{I_p}$ is only defined for $p\ge 1$, $\phi^{I_p}$ for $p\ge 0$, and $F^{I_P}$ for $p\ge -1$ with the additional constraint that $F^{I_{-1}}$ is $q$-exact.  Then the necessary condition for gauge invariance is that
\be
\label{eq:NecessaryAbelianCondition}
\lp q\al\rp\lp\pi_\La\otimes\pi_F\otimes\cdots\otimes\pi_F\rp=0.
\ee
This is really just a technical detail.  If we are given $V_p$ for $p\ge 0$, and we define $V_{-1}=q(V_0)$, by restriction from the full $V_{-1}$ if necessary, then given $\al$ satisfying (\ref{eq:NecessaryAbelianCondition}), we can always extend the definition of $\al$ to a new $\widetilde{\al}$ such that $q\widetilde{\al}=0$ and
\be
\widetilde{\al}\lp\pi_\phi\otimes\pi_F\otimes\cdots\otimes\pi_F\rp=\al\lp\pi_\phi\otimes\pi_F\otimes\cdots\otimes\pi_F\rp,
\ee
so that $S_{CS,\widetilde{\al}}=S_{CS,\al}$.  For this reason, we can often treat $q\al=0$ as both necessary and sufficient\footnote{There is an interesting reformulation that can be made here.  Suppose we consider a more general spacetime manifold $S$ which can be written as the boundary of some $(d+1)$-dimensional manifold $T$, and formally lift all of our fields to differential forms on $T$.  Then we can show that
\be
d\al(\phi,F,\cdots,F)=\al(F,F,\cdots,F)+\lp q\al\rp(\phi,F,\cdots,F).
\ee
If $q\al=0$, it then follows that
\be
S_{CS,\al}=\int_S\al(\phi,F,\cdots,F)=\int_T\al(F,F,\cdots,F).
\ee
It would be quite interesting to push this idea further for topologically interesting spaces, etc.  We would like to thank the JHEP referee for this suggestion.}.

What about exact $\al=q\beta$?  A calculation similar to the above shows in this case that
\be
S_{CS,q_{\widehat{X}}\beta}=-\int\beta(F,F,\cdots,F).
\ee
In other words, if $\al$ is $q$-exact, then $S_{CS,\al}$ can be constructed in terms of field strengths only.  If we are interested in Chern-Simons actions that can't be constructed from field strengths alone, then we should quotient out by the image of $q$.  This means that the gauge-invariant Chern-Simons actions are classified by the cohomology group
\be
H^{d+1-N}_q(\widehat{X}).
\ee

\subsection{Non-abelian Bosonic Chern-Simons Actions}

We take over all the structures from the abelian case, but for each $x\in\mathfrak{g}$ we have maps
\be
t_x:V_p\longrightarrow V_p,\qquad h_x:V_p\longrightarrow V_{p+1},
\ee
given by $t_x(\varphi)=t(x,\varphi)$ and $h_x(\varphi)=h(x,\varphi)$ respectively.  These can be lifted to maps on $X_{(N)}^\bullet$ or $\widehat{X}_{(N)}^\bullet$ by taking\footnote{When we promote everything to forms on $\R^d$ these expressions mostly work the same unless $x$ is an odd-degree form in space-time.  In that case, we need to introduce extra signs in these expressions for commuting $x$ through the $\varphi_i$.}, for $\al\in X_{(N)}^p$,
\be
\lp t_x\al\rp(\varphi_1,\cdots,\varphi_N)=\sum_{i=1}^N\al(\varphi_1,\cdots,t_x(\varphi_i),\cdots,\varphi_N),
\ee
and
\be
\lp h_x\al\rp(\varphi_1,\cdots,\varphi_N)=\sum_{i=1}^N\lp -1\rp^{\sum_{j=1}^{i-1}\deg(\varphi_j)+i+1}\al(\varphi_1,\cdots,h_x(\varphi_j),\cdots,\varphi_N).
\ee
With these definitions we have
\be
t_x:X_{(N)}^p\longrightarrow X_{(N)}^p,\qquad h_x:X_{(N)}^p\longrightarrow X_{(N)}^{p-1},
\ee
or the corresponding maps with $X_{(N)}$ replaced by $\widehat{X}_{(N)}$ by simple restriction (in other words these maps commute with projection or inclusion between the hatted and un-hatted complexes).

Just in the same way that $q^2=0$ on $V_\bullet$ implied that $q^2=0$ on $X_{(N)}^\bullet$ or $\widehat{X}^\bullet_{(N)}$, we can check that the various relations between the maps lift to the maps defined on $X_{(N)}^\bullet$.  Explicitly,
\begin{subequations}
\begin{align}
t_xt_y-t_yt_x=& t_{[x,y]},\\
t_xq-qt_x=& 0,\\
h_xt_y-t_yh_x=& h_{[x,y]},\\
\label{seq:qh}
qh_x+h_xq=& -t_x,\\
h_xh_y+h_yh_x=& 0,
\end{align}
\end{subequations}
and similarly for $\widehat{X}^\bullet$.

Turning to the hierarchy, we recall that the variations of the potentials become
\be
\d\phi=t_\la\phi+d\La-t_\mcA\La+q(\La)+h_\mcF\La,
\ee
and the Bianchi identities are
\be
dF-t_\mcA F=-q(F)-h_\mcF F.
\ee
Recall also that the field strengths are covariant,
\be
\d F=t_\la F.
\ee

As before, take $\al\in\widehat{X}_{(N)}^{d+1-N}$ and define\footnote{We could have tried something more general here, allowing Chern-Simons actions which depend explicitly on $\cal{A}$ and $\mcF$ as well.  In the present work we will neglect this possibility since the gauge variations always preserve the number of matter fields and increasing the number of non-abelian gauge fields appearing in the Chern-Simons action (either via field strengths $\mcF$ or having a potential $\cal{A}$ which displaces one of the potentials $\phi$ into a field strength $F$) will necessarily increase the dimension of the action. Also, the generalization does not appear in our motivating examples coming from dimensional reduction.}
\be
S_{CS,\al}=\int\al(\phi,F,\cdots,F).
\ee
Then under a $\la$ transformation we have
\begin{align}
\d_\la S_{CS,\al}=& \int\left\{\al(t_\la\phi,F,\cdots,F)+\al(\phi,t_\la F,F,\cdots,F)+\cdots+\al(\phi,F,\cdots,F,t_\la F)\right\}\non\\
=&\int \lp t_\la\al\rp(\phi,F,\cdots,F).
\end{align}
Under a $\La$ transformation, we have, after performing the now-familiar manipulations,
\begin{align}
\d_\La S_{CS,\al}=& \int\al(d\La-t_\mcA\La+q(\La)+h_\mcF\La,F,\cdots,F)\non\\
=& \int\left\{\lp q\al\rp(\La,F,\cdots,F)+\lp h_\mcF\al\rp(\La,F,\cdots,F)-\lp t_\mcA\al\rp(\La,F,\cdots,F)\right\}
\end{align}
From these expressions we see that a sufficient condition for gauge invariance is that
\bea
\label{eq:alphaProjections}
\lp q\al\rp\lp\pi_\La\otimes\pi_F\otimes\cdots\otimes\pi_F\rp &=& 0,\non\\
\lp h_x\al\rp\lp\pi_\La\otimes\pi_F\otimes\cdots\otimes\pi_F\rp &=& 0,\qquad\forall x\in\mathfrak{g}.
\eea
Note that these conditions imply also that
\be
\lp t_x\al\rp\lp\pi_\La\otimes\pi_F\otimes\cdots\otimes\pi_F\rp=0,\qquad \forall x\in\mathfrak{g},
\ee  
so the only extra condition we need to impose is (since $\pi_\phi=\pi_\La+\pi_0$) that
\be
\lp t_x\al\rp\lp\pi_0\otimes\pi_F\otimes\cdots\otimes\pi_F\rp=0,\qquad \forall x\in\mathfrak{g}.
\ee

As before, this is more or less a technicality, and in practice we can consider the condition for gauge invariance to be simply that $q\al=0$ and $h_x\al=0$, $\forall x\in\mathfrak{g}$.

\subsection{Explicit Equations for the Coefficients}

Specializing to $d=4$, we can expand the $\al$'s out and explicitly write the conditions for gauge invariance.  For instance the invariance of the linear Chern-Simons action,
\be
\label{eq:LinCSGI}
S_{CS,\al}=\int\al_XD^X_{[4]},
\ee
becomes simply that (note that $\al$ is automatically $q$-closed in this case)
\be
\al_X\lp h_k\rp^X_{\hph{X}S}=0.
\ee
The subscript on the potential $D^X_{[4]}$ simply indicates that it is a four-form in spacetime.

For a quadratic Chern-Simons action,
\begin{align}
\label{eq:QuadCSBosonicAction}
S_{CS,\al}=& \int\left\{\al_{1AS}a^A_{[0]}F^S_{[4]}+\al_{2IM}A^I_{[1]}\w F^M_{[3]}+\al_{3MI}B^M_{[2]}\w F^I_{[2]}\right.\non\\
& \qquad\left. +\al_{4SA}C^S_{[3]}\w F^A_{[1]}+\al_{5XZ}D^X_{[4]}F^Z_{[0]}\right\}.
\end{align}
our conditions are
\begin{subequations}
\label{eq:QuadCSalphaq}
\begin{align}
\al_{1AS}q^A_{\hph{A}I}+\al_{2IM}q^M_{\hph{M}S}=& 0,\\
\al_{2IN}q^I_{\hph{I}M}-\al_{3MI}q^I_{\hph{I}N}=& 0,\\
\al_{3MI}q^M_{\hph{M}S}+\al_{4SA}q^A_{\hph{A}I}=& 0,\\
\al_{4SA}q^S_{\hph{S}X}-\al_{5XZ}q^Z_{\hph{Z}A}=& 0,
\end{align}
\end{subequations}
from $(q\al)(\pi_\La\otimes\pi_F)=0$,
\begin{subequations}
\label{eq:QuadCSalphah}
\begin{align}
\al_{2IM}\lp h_k\rp^M_{\hph{M}J}+\al_{3MJ}\lp h_k\rp^M_{\hph{M}I}=& 0,\\
-\al_{3MI}\lp h_k\rp^I_{\hph{I}A}+\al_{4SA}\lp h_k\rp^S_{\hph{S}M}=& 0,\\
\al_{4SB}\lp h_k\rp^B_{\hph{A}Z}q^Z_{\hph{Z}A}+\al_{5XZ}\lp h_k\rp^X_{\hph{X}S}q^Z_{\hph{Z}A}=& 0,
\end{align}
\end{subequations}
from $(h_x\al)(\pi_\La\otimes\pi_F)=0$, and
\be
\label{eq:QuadCSalphat}
\al_{1BS}\lp t_k\rp^B_{\hph{B}A}+\al_{1AT}\lp t_k\rp^T_{\hph{T}S}=0,
\ee
from $(t_x\al)(\pi_0\otimes\pi_F)=0$.

For a cubic Chern-Simons action, there are nine coefficients appearing in $\al$, 
\be
\left\{\al_{1AZS},\al_{2ABM},\al_{3A(IJ)},\al_{4IZM},\al_{5IAJ},\al_{6MZI},\al_{7M[AB]},\al_{8SZA},\al_{9X(ZZ')}\right\},
\ee
where we have noted where they are symmetric or antisymmetric.  The conditions they must satisfy are
\begin{subequations}
\label{eq:CubicCSalphaq}
\begin{align}
\al_{1BZS}q^B_{\hph{B}I}q^Z_{\hph{Z}A}+\al_{4IZM}q^M_{\hph{M}S}q^Z_{\hph{Z}A}=& 0,\\
\al_{2BAM}q^B_{\hph{B}I}+\al_{4IZM}q^Z_{\hph{Z}A}-\al_{5IAJ}q^J_{\hph{J}M}=& 0,\\
\al_{3AJK}q^A_{\hph{A}I}+\al_{5IA(J}q^A_{\hph{A}K)}=& 0,\\
\al_{4IZN}q^I_{\hph{I}M}q^Z_{\hph{Z}A}-\al_{6MZI}q^I_{\hph{I}N}q^Z_{\hph{Z}A}=& 0,\\
\al_{5JAI}q^J_{\hph{J}M}-\al_{6MZI}q^Z_{\hph{Z}A}+2\al_{7MAB}q^B_{\hph{B}I}=& 0,\\
\al_{6MZI}q^M_{\hph{M}S}q^Z_{\hph{Z}A}+\al_{8SZB}q^B_{\hph{B}I}q^Z_{\hph{Z}A}=& 0,\\
\al_{7MAB}q^M_{\hph{M}S}-\al_{8SZ[A}q^Z_{\hph{Z}B]}=& 0,\\
\al_{8SZA}q^S_{\hph{S}X}q^Z_{\hph{Z}B}-2\al_{9XZZ'}q^Z_{\hph{Z}B}q^{Z'}_{\hph{Z'}A}=& 0,
\end{align}
\end{subequations}
\begin{subequations}
\label{eq:CubicCSalphah}
\begin{align}
\al_{4IZM}\lp h_k\rp^M_{\hph{M}J}q^Z_{\hph{Z}A}+\al_{5IBJ}\lp h_k\rp^B_{\hph{B}Z}q^Z_{\hph{Z}A}+\al_{6MZJ}\lp h_k\rp^M_{\hph{M}I}q^Z_{\hph{Z}A}=& 0,\\
-\al_{5I[A|J|}\lp h_k\rp^J_{\hph{J}B]}+\al_{7MAB}\lp h_k\rp^M_{\hph{M}I}=& 0,\\
-\al_{6MZI}\lp h_k\rp^I_{\hph{I}A}q^Z_{\hph{Z}B}-2\al_{7MCA}\lp h_k\rp^C_{\hph{C}Z}q^Z_{\hph{Z}B}+\al_{8SZA}\lp h_k\rp^S_{\hph{S}M}q^Z_{\hph{Z}B}=& 0,\\
\al_{8SZC}\lp h_k\rp^C_{\hph{C}Z'}q^Z_{\hph{Z}(A}q^{Z'}_{\hph{Z'}B)}+\al_{9XZZ'}\lp h_k\rp^X_{\hph{X}S}q^Z_{\hph{Z}(A}q^{Z'}_{\hph{Z'}B)}=& 0,
\end{align}
\end{subequations}
and
\begin{subequations}
\label{eq:CubicCSalphat}
\begin{align}
\al_{1CZS}\lp t_k\rp^C_{\hph{C}A}q^Z_{\hph{Z}B}+\al_{1AZ'S}\lp t_k\rp^{Z'}_{\hph{Z'}Z}q^Z_{\hph{Z}B}+\al_{1AZT}\lp t_k\rp^T_{\hph{T}S}q^Z_{\hph{Z}B}=& 0,\\
\al_{2CBM}\lp t_k\rp^C_{\hph{C}A}+\al_{2ACM}\lp t_k\rp^C_{\hph{C}B}+\al_{2ABN}\lp t_k\rp^N_{\hph{N}M}=& 0,\\
\al_{3BIJ}\lp t_k\rp^B_{\hph{B}A}+2\al_{3AK(I}\lp t_k\rp^K_{\hph{K}J)}=& 0.
\end{align}
\end{subequations}
We will provide an explicit solution to these equations in section \ref{sec:DimRed}.

\section{Superfield Chern-Simons Actions}
\label{sec:SCSAs}
Now we would like to supersymmetrize the structures we found in section \ref{sec:BCSAs} to $N=1$ superspace in four dimensions.  Our starting point will be the abelian Chern-Simons actions that we constructed in ref.\ \cite{Becker:2016xgv}, but where we promote all derivatives to covariant derivatives, and use the field strengths constructed in (\ref{eqs:FieldStrengths}).  When expanded in components, these actions will contain the bosonic Chern-Simons actions of section \ref{sec:BCSAs} (along with more pieces involving other component fields), and are gauge-invariant when we restrict to the abelian case.  

It remains to check that they remain invariant in the non-abelian case.  For the non-abelian gauge variations with parameter $\la$, it will be easy to see that the condition for gauge invariance will simply be that $t_\la\al=0$, just as in the bosonic case, and since the action of $t_\la$ preserves $V_\bullet$ degree, this means that $t_x$ annihilates the Lagrangian term by term.  Once this is established, it is possible to go back and forth between full superspace integrals and chiral superspace integrals using covariant derivatives,
\be
d^2\th\sim -\frac{1}{4}\mcD^2,\qquad d^2\bar{\th}\sim -\frac{1}{4}\ov{\mcD}^2,\qquad d^4\th\sim\frac{1}{16}\mcD^2\ov{\mcD}^2.
\ee
We will still need to check that the actions are invariant under the hierarchy gauge transformations, and in fact we will find a surprise in the case of the cubic Chern-Simons action, where an additional piece will have to be added to make the action fully gauge-invariant.

\subsection{Linear Super-Chern-Simons Action}

We start with the linear Chern-Simons action
\be
S_{0,SCS}=\operatorname{Re}\ls i\int d^4xd^2\th\,\al(\G)\rs,
\ee
where we use the short-hand $\al(\G)=\al_X\G^X$.

Under the variations (\ref{eqs:SuperfieldVariations}), we have
\begin{align}
\d S_{0,SCS}=& \operatorname{Re}\ls i\int d^4xd^2\th\,\al(t(\la,\G)-\frac{1}{4}\ov{\mcD}^2\Xi+h(\mcW^\al,\Upsilon_\al))\rs\non\\
=& \operatorname{Re}\ls i\int d^4xd^2\th\lp\lp t_\la\al\rp(\G)+\lp h_{\mcW^\al}\al\rp(\Upsilon_\al)\rp\rs.
\end{align}
The $\Xi$ term vanishes since we can promote it to $\operatorname{Re}[i\int d^4xd^4\th\Xi]=0$, since $\Xi$ is real.  Moreover, the condition that $h_x\al=0$ implies that $t_x\al=0$ in this case (using (\ref{seq:qh}) and $q\al=0$), so the only condition for gauge invariance is eqn.\ (\ref{eq:LinCSGI}), just as in the bosonic case.

\subsection{Quadratic Super-Chern-Simons Action}

For the quadratic Chern-Simons term, we have
\begin{align}
S_{1,SCS}=& \int d^4xd^4\th\lp\al_2(V,H)-\al_4(X,F)\rp\non\\
& \qquad +\operatorname{Re}\ls i\int d^4xd^2\th\lp\al_1(\Phi,G)+\al_3(\Sigma^\al,W_\al)+\al_5(\G,E)\rp\rs.
\end{align}
The $\al$'s are as in (\ref{eq:QuadCSBosonicAction}), in notation which is hopefully obvious (i.e.\ $\al_1(\Phi,G)=\al_{1AS}\Phi^AG^S$, etc.).

Under the non-abelian variation we simply get the condition $t_\la\al=0$.  Now consider the other variations.  After some algebraic manipulations involving integrations by parts, the algebra of super-covariant derivatives, and the Bianchi identities on the field strengths, we find
\begin{subequations}
\begin{align}
\d_\La S_{1,SCS}=& \operatorname{Re}\ls i\int d^4xd^2\th\lp\al_1(q(\La),G)+\al_2(\La,q(G))\right.\right.\non\\
& \qquad\left.\left. +\al_2(\La,h_{\mcW^\al}W_\al)+\al_3(h_{\mcW^\al}\La,W_\al)\rp\vphantom{\int}\rs,\\
\d_US_{1,SCS}=& \int d^4xd^4\th\lp\al_2(q(U),H)-\al_3(U,q(H))\right.\non\\
& \qquad\left. -\al_4(\Om_h(\mcW,U),F)-\al_3(U,\Om_h(\mcW,F))\rp,\\
\d_\Upsilon S_{1,SCS}=& \operatorname{Re}\ls i\int d^4xd^2\th\lp\al_3(q(\Upsilon^\al),W_\al)+\al_4(\Upsilon^\al,q(W_\al))\right.\right.\non\\
& \qquad\left.\left.+\al_4(\Upsilon^\al,h_{\mcW_\al}E)+\al_5(h_{\mcW^\al}\Upsilon_\al,E)\rp\vphantom{\int}\rs,\\
\d_\Xi S_{1,SCS}=& \int d^4xd^4\th\lp -\al_4(q(\Xi),F)+\al_5(\Xi,q(F))\rp.
\end{align}
\end{subequations}
It is easy to confirm that the vanishing of these variations is precisely equivalent to the conditions (\ref{eq:QuadCSalphaq}), (\ref{eq:QuadCSalphah}), and (\ref{eq:QuadCSalphat}) of the bosonic case.

\subsection{Cubic Super-Chern-Simons Action}

In order to write the cubic super-Chern-Simons action from ref.\ \cite{Becker:2016xgv}, and take its variations, we need to make a couple of definitions,
\be
\widehat{\Phi}=\hlf\lp\Phi+\ov{\Phi}\rp,\qquad\widehat{E}=\hlf\lp E+\ov{E}\rp=-q(\widehat{\Phi}),\qquad\widehat{\La}=\hlf\lp\La+\ov{\La}\rp,
\ee
and another application of the Chern-Simons superfield construction, where we are given three arguments, one of which is a chiral spinor superfield $\psi$ and the other two are real scalar superfields $U_1$ and $U_2$.  Then we have
\begin{subequations}
\begin{align}
\Om_\al(\psi,U_1,U_2)=& \al(\psi^\al,U_1,\mcD_\al U_2)+\al(\ov{\psi}_{\dot\al},U_1,\ov{\mcD}^{\dot\al}U_2)+\hlf\al(\mcD^\al\psi_\al+\ov{\mcD}_{\dot\al}\ov{\psi}^{\dot\al},U_1,U_2),\\
\Om_\al(U_2,\psi,U_1)=& \al(\mcD^\al U_2,\psi_\al,U_1)+\al(\ov{\mcD}_{\dot\al}U_2,\ov{\psi}^{\dot\al},U_1)+\hlf\al(U_2,\mcD^\al\psi_\al+\ov{\mcD}_{\dot\al}\ov{\psi}^{\dot\al},U_1),\\
\Om_\al(U_1,U_2,\psi)=& \al(U_1,\mcD^\al U_2,\psi_\al)+\al(U_1,\ov{\mcD}_{\dot\al}U_2,\ov{\psi}^{\dot\al})+\hlf\al(U_1,U_2,\mcD^\al\psi_\al+\ov{\mcD}_{\dot\al}\ov{\psi}^{\dot\al}).
\end{align}
\end{subequations}
As examples
\begin{subequations}
\begin{align}
\Om_{\al_5}(V,F,W)=& \al_5(V,\mcD^\al F,W_\al)+\al_5(V,\ov{\mcD}_{\dot\al}F,\ov{W}^{\dot\al})+\hlf\al_5(V,F,\mcD^\al W_\al+\ov{\mcD}_{\dot\al}\ov{W}^{\dot\al}),\\
\Om_{\al_7}(\Sigma,F,F)=& \al_7(\Sigma^\al,F,\mcD_\al F)+\al_7(\ov{\Sigma}_{\dot\al},F,\ov{\mcD}^{\dot\al}F),
\end{align}
\end{subequations}
Note that the last term in $\Om_{\al_7}$ vanishes since
\be
\al_7(\mcD^\al\Sigma_\al+\ov{\mcD}_{\dot\al}\ov{\Sigma}^{\dot\al},F,F)=0,
\ee
by the antisymmetry of $\al_7$ in its last two arguments.

With these definitions, we have the cubic Chern-Simons action from ref.~\cite{Becker:2016xgv}, which was invariant in the abelian case and correctly reproduced the bosonic Chern-Simons action (\ref{eq:CS2Indices}),
\begin{align}
S^{(0)}_{2,SCS}=& \int d^4xd^4\th\lp\al_2(\widehat{\Phi},F,H)+\al_4(V,\widehat{E},H)+\Om_{\al_5}(V,F,W)+\Om_{\al_7}(\Sigma,F,F)\right.\non\\
& \qquad\left. -\al_8(X,\widehat{E},F)\rp+\operatorname{Re}\ls i\int d^4xd^2\th\lp\al_1(\Phi,E,G)+\al_3(\Phi,W^\al,W_\al)\right.\right.\non\\
& \qquad\left.\left. +\al_6(\Sigma^\al,E,W_\al)+\al_9(\G,E,E)\rp\vphantom{\int}\rs.
\end{align}
The superscript $(0)$ is because we will find that a correction will need to be added to get a gauge-invariant action.

Again, it is easy to check that the $\la$ variation simply leads to the condition that $t_\la\al=0$.  For the others, we have (after significant algebra)
\begin{subequations}
\begin{align}
\d_\La S^{(0)}_{2,SCS}=& \int d^4xd^4\th\left\{\vphantom{\frac{\La-\ov{\La}}{2i}}\lp\al_2(q(\widehat{\La}),F,H)+\al_4(\widehat{\La},q(F),H)-\al_5(\widehat{\La},F,q(H))\rp\right.\non\\
& \left. -\operatorname{Re}\ls\al_5(\La,F,h_{\mcW^\al}\mcD_\al F)-\al_5(\La,\mcD^\al F,h_{\mcW_\al}F)-2\al_7(h_{\mcW^\al}\La,F,\mcD_\al F)\rs\right.\non\\
& \qquad\left. -i\al_5(\frac{\La-\ov{\La}}{2i},\mcD^\al F,h_{\mcW_\al}F)+i\al_5(\frac{\La-\ov{\La}}{2i},\ov{\mcD}_{\dot\al}F,h_{\ov{\mcW}^{\dot\al}}F)\right\}\non\\
& \quad +\operatorname{Re}\ls i\int d^4xd^2\th\left\{\lp\al_1(q(\La),E,G)+\al_4(\La,E,q(G))\rp\right.\right.\non\\
& \qquad\left.\left. +\lp\al_3(q(\La),W^\al,W_\al)+\al_5(\La,q(W^\al),W_\al)\rp\right.\right.\non\\
& \qquad\left.\left. +\lp\al_4(\La,E,h_{\mcW^\al}W_\al)+\al_5(\La,h_{\mcW^\al}E,W_\al)+\al_6(h_{\mcW^\al}\La,E,W_\al)\rp\right\}\vphantom{\int}\rs,
\end{align}
\begin{align}
\d_US^{(0)}_{2,SCS}=& \int d^4xd^4\th\left\{\lp\al_4(q(U),\widehat{E},H)-\al_6(U,\widehat{E},q(H))\rp\right.\non\\
& \quad\left. +\lp\Om_{\al_5}(q(U),F,W)-\Om_{\al_6}(U,q(F),W)+2\Om_{\al_7}(U,F,q(W))\rp\right.\non\\
& \quad\left. +\lp -\al_6(U,\widehat{E},h_{\mcD\mcW}F)-2\al_7(U,h_{\mcD\mcW}\widehat{E},F)+\al_8(h_{\mcD\mcW}U,\widehat{E},F)\rp\right.\non\\
& \quad\left. +2\operatorname{Re}\ls -\al_6(U,\widehat{E},h_{\mcW^\al}\mcD_\al F)-2\al_7(U,h_{\mcW^\al}\widehat{E},\mcD_\al F)\right.\right.\non\\
& \qquad\left.\left. +\al_8(h_{\mcW^\al}U,\widehat{E},\mcD_\al F)\rs +2\operatorname{Re}\ls i\al_6(U,q(\mcD^\al F),h_{\mcW_\al}F)\right.\right.\non\\
& \qquad\left.\left. +2i\al_7(U,h_{\mcW^\al}q(\mcD_\al F),F)-i\al_8(h_{\mcW^\al}U,q(\mcD_\al F),F)\rs\right.\non\\
& \quad\left. +2\operatorname{Re}\ls i\al_5(q(U),\mcD^\al F,h_{\mcW_\al}F)-i\al_6(U,q(\mcD^\al F),h_{\mcW_\al}F)\right.\right.\non\\
& \qquad\left.\left. +2i\al_7(U,\mcD^\al F,q(h_{\mcW_\al}F))\rs\right.\non\\
& \quad\left. -i\al_5(q(U),\mcD^\al F,h_{\mcW_\al}F)+i\al_5(q(U),\ov{\mcD}_{\dot\al}F,h_{\ov{\mcW}^{\dot\al}}F)\vphantom{\widehat{E}}\right\},
\end{align}
\begin{align}
\d_\Upsilon S^{(0)}_{2,SCS}=& \int d^4xd^4\th\lp\Om_{\al_7}(q(\Upsilon),F,F)+\Om_{\al_8}(\Upsilon,q(F),F)\rp\non\\
& \quad +\operatorname{Re}\ls i\int d^4xd^2\th\left\{\lp\al_6(q(\Upsilon^\al),E,W_\al)+\al_8(\Upsilon^\al,E,q(W_\al))\rp\right.\right.\non\\
& \qquad\left.\left. +\lp\al_8(\Upsilon^\al,E,h_{\mcW_\al}E)+\al_9(h_{\mcW^\al}\Upsilon_\al,E,E)\rp\right\} \vphantom{\int}\rs,
\end{align}
\begin{align}
\d_\Xi S^{(0)}_{2,SCS}=& \int d^4xd^4\th\lp -\al_8(q(\Xi),\widehat{E},F)+2\al_9(\Xi,\widehat{E},q(F))\rp.
\end{align}
\end{subequations}
From these expressions, we can see that invariance requires precisely the same conditions (\ref{eq:CubicCSalphaq}) and (\ref{eq:CubicCSalphah}) as in the bosonic case, but even after imposing these conditions, the action is not completely invariant; we have a remainder term
\be
\d S^{(0)}_{2,SCS}=-\operatorname{Re}\ls i\int d^4xd^4\th\,\al_5(\frac{\La-\ov{\La}}{2i}+q(U),\mcD^\al F,h_{\mcW_\al}F)\rs.
\ee
Written in this form, it's obvious that we can cancel the variation, by adding the additional piece $i\int d^4xd^4\th\,\al_5(V,\mcD^\al F,h_{\mcW_\al}F)$ to the action. Doing so, we arrive at the final form
\begin{align}
\label{eq:FinalCubicCS}
S_{2,SCS}=&
\int d^4xd^4\th\lp\al_2(\widehat{\Phi},F,H)+\al_4(V,\widehat{E},H)+\Om_{\al_5}(V,F,W)+\Om_{\al_7}(\Sigma,F,F)\right.\non\\
& \qquad\left. -\al_8(X,\widehat{E},F)
+\operatorname{Re}\ls i\al_5(V,\mcD^\al F,h_{\mcW_\al}F)\rs\rp
\non\\
& \quad +\operatorname{Re}\ls i\int d^4xd^2\th\lp\al_1(\Phi,E,G)+\al_3(\Phi,W^\al,W_\al)\right.\right.\non\\
& \qquad\left.\left. +\al_6(\Sigma^\al,E,W_\al)+\al_9(\G,E,E)\rp\vphantom{\int}\rs
\end{align}
with the $\alpha$s satisfying eqns. (\ref{eq:CubicCSalphaq}), (\ref{eq:CubicCSalphah}), and (\ref{eq:CubicCSalphat}).

\section{Dimensional Reduction}
\label{sec:DimRed}

One of the prime motivations for this work is to develop the machinery needed to describe a higher-dimensional supergravity theory, for example eleven-dimensional supergravity, in an off-shell four-dimensional $N=1$ formulation.\footnote{This is analogous to the construction ten-dimensional super-Yang-Mills theory in terms of 4D, $N=1$ superfield representations \cite{Marcus:1983wb}.} In particular, when reducing a $p$-form gauge potential, such as the three-form in eleven-dimensions, one naturally encounters hierarchies of the sort described in this paper.  The matter fields, in $V_\bullet$, arise from reductions of the $p$-form itself, while the non-abelian gauge field is the Kaluza-Klein vector, and the corresponding gauge group is the group of diffeomorphisms of the internal space $M$, with $\mathfrak{g}\cong TM$.  Let us now make these observations more precise.

\subsection{Hierarchy from Reduction}

As described in section \ref{intprod}, if we reduce a theory with an $n$-form potential in $D$ dimensions down to $d$ dimensions on a $(D-d)$-dimensional manifold $M$, we are in the situation described by our tensor hierarchy.  We have
\be
V_p\cong\Om^{n-p}(M),
\ee
with bases labeled by multi-indices
\be
I_p=(i_1\cdots i_{n-p};y),
\ee
where $i_k$ are indices on $M$ and $y$ is a coordinate on $M$.  We will use somewhat interchangeably the following,
\be
\varphi^{I_p}=\varphi_{(i_1\cdots i_{n-p};y)}=\varphi_{i_1\cdots i_{n-p}}(y).
\ee
Note that when we write the indices $i_j$ out explicitly, it is natural to put them downstairs since they correspond to differential forms on $M$.  A summation over a repeated index involves both a standard summation over the $i_j$ indices, as well as an integration of $y$ over $M$.  Simlarly, we will use
\be
\bar{k}=(k;y),\qquad x^{\bar{k}}=x^{(k;y)}=x^k(y),
\ee
for indices of $\mathfrak{g}\cong TM$.

In this language,
\be
f^{(k;y)}_{\hph{(k;y)}(\ell;y')(m;y'')}=-\d^k_\ell\lp\pa_m\d\rp(y-y')\d(y-y'')+\d^k_m\d(y-y')\lp\pa_\ell\d\rp(y-y''),
\ee
\be
q_{(i_1\cdots i_{n-p};y)}^{\hph{(i_1\cdots i_{n-p};y)}(j_1\cdots j_{n-p-1};y')}=\lp -1\rp^{n-1}\lp n-p\rp\d_{[i_1}^{[j_1}\cdots\d_{i_{n-p-1}}^{j_{n-p-1}]}\lp\pa_{i_{n-p}]}\d\rp(y-y'),
\ee
\be
\lp h_{(k;u)}\rp_{(i_1\cdots i_{n-p};y)}^{\hph{(i_1\cdots i_{n-p};y)}(j_1\cdots j_{n-p+1};y')}=\lp -1\rp^p\d_k^{[j_1}\d_{[i_1}^{j_2}\cdots\d_{i_{n-p}]}^{j_{n-p+1}]}\d(u-y)\d(u-y').
\ee
and
\begin{multline}
\lp t_{(k;u)}\rp_{(i_1\cdots i_{n-p};y)}^{\hph{(i_1\cdots i_{n-p};y)}(j_1\cdots j_{n-p};y')}=\d_{[i_1}^{[j_1}\cdots\d_{i_{n-p}]}^{j_{n-p}]}\lp\pa_k\d\rp(u-y')\d(u-y)\\
+\lp -1\rp^{n-p}\lp n-p\rp\d_k^{[j_1}\d_{[i_1}^{j_2}\cdots\d_{i_{n-p-1}}^{j_{n-p}]}\lp\pa_{i_{n-p}]}\d\rp(u-y)\d(y-y').
\end{multline}
We are using the notation that $(\pa\d)$ is the derivative of the delta function with respect to its argument, so for example
\be
\lp\pa_k\d\rp(u-y')=\frac{\pa}{\pa u^k}\ls\d(u-y')\rs=-\frac{\pa}{\pa y^{\prime\,k}}\ls\d(u-y')\rs.
\ee

One can check explicitly that these coefficients satisfy the required conditions, but its easier to see by computing their action on fields.  For example, we have
\be
f^{(k;y)}_{\hph{(k;y)}(\ell;y')(m;y'')}x_1^{(\ell;y')}x_2^{(m;y'')}=-x_2^\ell(y)\pa_\ell x_1^k(y)+x_1^\ell(y)\pa_\ell x_2^k(y),
\ee
which is simply the Lie bracket on vector fields, and it is easy to check antisymmetry and the Jacobi identity.

Similarly,
\be
q_{(i_1\cdots i_{n-p};y)}^{\hph{(i_1\cdots i_{n-p};y)}(j_1\cdots j_{n-p-1};y')}\varphi_{(j_1\cdots j_{n-p-1};y')}=\lp -1\rp^p\lp n-p\rp\pa_{[i_1}\varphi_{i_2\cdots i_{n-p}]}(y),
\ee
or
\be
q(\varphi)=\lp -1\rp^pd_M\varphi,
\ee
where $d_M$ is the exterior derivative acting on forms on $M$.

Next,
\be
\lp h_{(k;u)}\rp_{(i_1\cdots i_{n-p};y)}^{\hph{(i_1\cdots i_{n-p};y)}(j_1\cdots j_{n-p+1};y')}x^{(k;u)}\varphi_{(j_1\cdots j_{n-p+1};y')}=\lp -1\rp^px^k(y)\varphi_{ki_1\cdots i_{n-p}}(y),
\ee
or
\be
h_x\varphi=\lp -1\rp^p\iota_x\varphi,
\ee
contraction with the vector $x\in TM$, and
\begin{multline}
\lp t_{(k;u)}\rp_{(i_1\cdots i_{n-p};y)}^{\hph{(i_1\cdots i_{n-p};y)}(j_1\cdots j_{n-p};y')}x^{(k;u)}\varphi_{(j_1\cdots j_{n-p};y')}\\
=x^k(y)\pa_k\varphi_{i_1\cdots i_{n-p}}(y)+\lp n-p\rp\pa_{[i_1}x^k(y)\varphi_{|k|i_2\cdots i_{n-p}]}(y),
\end{multline}
i.e.\ 
\be
t_x\varphi=\mathcal{L}_x\varphi,
\ee
the Lie derivative along the vector $x$.

Using this language, the relations among $f$, $q$, $h$, and $t$ are simply the equations stated already in section \ref{intprod},
\begin{subequations}
\begin{align}
\mathcal{L}_x\mathcal{L}_y-\mathcal{L}_y\mathcal{L}_x=& \mathcal{L}_{[x,y]},\\
d_M\mathcal{L}_x-\mathcal{L}_xd_M=& 0,\\
d_M^2=& 0,\\
\iota_x\mathcal{L}_y-\mathcal{L}_y\iota_x=& \iota_{[x,y]},\\
d_M\iota_x+\iota_xd_M=& \mathcal{L}_x,\\
\iota_x\iota_y+\iota_y\iota_x=& 0.
\end{align}
\end{subequations}

\subsection{Chern-Simons Actions}

Now suppose the $D$-dimensional theory has a Chern-Simons action.  For example, the eleven-dimensional supergravity theory contains a coupling
\be
S_{11D,CS}=\int C_{[3]}\w G_{[4]}\w G_{[4]},
\ee
where $G_{[4]}=dC_{[3]}$.  In general a theory with a single $n$-form potential can have a Chern-Simons action with $N-1$ field strengths if the total dimension of spacetime is $D=Nn+N-1$.  If $N>2$, we also need $n$ to be odd, otherwise the wedge product of field strengths will be zero automatically (if $N=2$ we should also have $n$ odd, otherwise the Chern-Simons term is a total derivative).  Our example above has $N=3$, $n=3$, $D=11$, but we can also have $N=3$, $n=1$, $D=5$, or other combinations.

In eqn.~(4.29) of ref.~\cite{Becker:2016xgv}, we gave a collection of coefficients $\al$, corresponding to the dimensional reduction of the eleven-dimensional Chern-Simons term to four dimensions, that satisfied the conditions (\ref{eq:CubicCSalphaq}) for gauge invariance of the abelian action.  It is possible to check that these same $\al$'s also satisfy the remaining conditions (\ref{eq:CubicCSalphah}) and (\ref{eq:CubicCSalphat}) of the non-abelian case.  It is not true that the $\al\in X_{(3)}^\bullet$ built from these coefficients satisfies $q\al=0$ or $h_x\al=0$; these conditions only hold after applying the additional projectors as in (\ref{eq:alphaProjections}).  However, by adding more coefficients to $\alpha$ which do not contribute to the action (since they are annihilated by the projectors $\pi_\phi$ or $\pi_F$), we can build an explicit $\alpha$ which is annihilated by $q$ and $h_x$.  This new $\alpha$ has a very nice interpretation of simply wedging together to get a top form which is then integrated over the internal space.

Indeed, upon reduction to $d$ dimensions, the Chern-Simons action will become a sum of terms of the form that we have described in section \ref{sec:BCSAs}.  The $\al$ in this case takes $N$ arguments that are forms on $M$ whose total degree is $D-d$, wedges them together to get a top form on $M$, and integrates the top form to get a number, i.e.\ 
\be
\al(\varphi_1,\cdots,\varphi_N)=\int_M\varphi_1\w\cdots\w\varphi_N.
\ee
Let us check that $q\al=0$ and $h_x\al=0$.  Well, $q\al$ will again take $N$ forms, now whose total degree is $D-d-1$, and a direct computation shows that
\be
\lp q\al\rp(\varphi_1,\cdots,\varphi_N)=\lp -1\rp^{\operatorname{deg}(\varphi_1)}\int_Md\lp\varphi_1\w\cdots\w\varphi_N\rp=0.
\ee
In other words, $q\al$ is zero because it is the integral of a total derivative.  $h_x\al=0$ for an even simpler reason, which is that $h_x\al$ would be the integral of the contraction of $x^k$ on a $(D-d+1)$-form on $M$.  But since there are no forms of degree greater than $D-d$ (the dimension of $M$), then this must be zero.
This shows that such an $\al$ indeed corresponds to a gauge-invariant Chern-Simons term (which should not come as a surprise).

Finally, note that the super-Chern-Simons actions, when expanded in component fields, will give rise to the bosonic Chern-Simons actions but also to many other terms involving other component fields.  Some of these additional terms can have nice interpretations.  For instance, the term given by $\al_3$ in (\ref{eq:FinalCubicCS}) will give rise to both a familiar axionic term
\be
\int aF\wedge F,
\ee
but also to a kinetic term (assuming that $\varphi$ gets a VEV)
\be
\int\varphi F\wedge\ast F.
\ee

\section{Prospects}
\label{sec:Prospects}
In this work, we have gauged the abelian superspace tensor hierarchy of reference \cite{Becker:2016xgv} by a non-abelian algebra $\mathfrak g$. In doing so, we have found that the required mathematical structure is that of a $\mathfrak g$-equivariant double complex of differential forms with values in representations of $\mathfrak g$. This action action is homotopically trivial and the homotopy operator is itself a differential. 
This gives an extension of the usual Lie derivative along $\mathfrak g$ vector fields to the complex of representations. 

Using these ingredients, we constructed manifestly supersymmetric actions including those of Chern-Simons type assuming certain cocycles exist on the tensor algebra of the total complex. Although the explicit equations defining the latter are somewhat imposing, existence of solutions is guaranteed by examples arising from decomposing higher-dimensional theories of $p$-forms in terms of four-dimensional representations. This was illustrated explicitly in section \ref{sec:DimRed} in the case of the eleven-dimensional gauge 3-form resulting in an embedding of this structure into a theory of superforms in 4D, $N=1$ superspace. 

Our eventual goal for this type of construction is to build a manifestly 4D, $N=1$ covariant description of eleven-dimensional supergravity. To that point, there are a few questions left unanswered at this stage of development. The most pressing of these is the following: the Chern-Simons action just constructed for eleven-dimensional supergravity is not eleven-dimensionally Lorentz invariant because there are component fields in the 4D, $N=1$ supermultiplets that are not present in the four-dimensional decomposition of the components of eleven-dimensional supergravity. 
Alternatively, since we have not included any of the 4D, $N=1$ supergravity fields, we have, at this stage, a non-gravitational theory partially encoding the structure of a purely gravitational one. 
The goal, then, is to couple the part of the theory we have just constructed to 4D, $N=1$ supergravity in just such a way that these two problems cancel.
A related problem is that the known on-shell descriptions of such dimensionally-reduced supergravity theories all require duality transformations on the component fields. How this is resolved is currently under investigation but precisely this question in the (very good) analogy of 5D, $N=1$ supergravity must have an answer given that the full off-shell structure of the latter is fully understood \cite{Banerjee:2011ts}. (See also refs.\ \cite{Linch:2002wg,Gates:2003qi} where this issue is addressed at the level of superfields.)

Along a very different line of investigation, our result also raises questions pertaining to related attempts to use similar non-abelian hierarchies for other purposes. In ref.\ \cite{Samtleben:2011fj} the original idea was to use such hierarchies to construct 6D, $N=(1,0)$ superconformal theories and, although the dimensions and supersymmetries differ, in retrospect our construction is morally the same. Furthermore, a moment of reflection suffices to conclude that the dimension and supersymmetry are largely irrelevant to the consistency of the basic hierarchy so it is natural to contemplate the relation between our results. Although the full exploration of this relationship is beyond the scope of this paper, we can already identify (at least two) interesting differences: The first is that the couplings studied here are of the same class as those arising from compactification and therefore {\it a priori} not as general as those considered in ref.\ \cite{Samtleben:2011fj}. 
On the other hand, in the approach of ref.\ \cite{Samtleben:2011fj} all vector fields (abelian and non-abelian) are treated on the same footing and the tensor analogous to the map (\ref{eq:h}) is naturally symmetric\footnote{More explicitly, once we combine $\mathfrak{g}$ and $V_1$ into a single vector space $\widehat{V}_1=V_1\oplus\mathfrak{g}$, then the analog of (\ref{eq:h}) for $r=1$ is
\be
\widehat{h}:(\Om^p\otimes\widehat{V}_1)\times(\Om^q\otimes\widehat{V}_1)\longrightarrow\Om^{p+q}\otimes V_2,
\ee
and it is sensible to ask about $\widehat{h}$ being symmetric or antisymmetric in its arguments.} in contrast to the asymmetric cases considered here. 

Finally, we should point out that the construction presented here is a hybrid of two approaches in which the forms in the hierarchy are treated in terms of superspace ``prepotentials'' whereas the non-abelian gauging is treated in terms of superspace potentials. Ultimately, the use of the prepotentials is what is to blame for the complexity of the analysis throughout this paper. Morally speaking, the entire analysis should be done without recourse to this pre-geometry. If this were possible, none of the complicated $D$-algebra should be needed and, similarly, no part of the analysis should require the myriad ``magical'' cancelations. In a forthcoming paper, we hope to show this concretely by recasting the results presented here in terms of the geometry of superforms \cite{BBLRR}.

\section*{Acknowledgements}
We thank 
Daniel Butter for discussions,
Robert Wimmer for correspondence elucidating certain aspect of the non-abelian tensor hierarchy, and
Stephen Randall for numerous discussions and collaboration on closely-related topics.
This work was supported by NSF grants
PHY-1214333 and PHY-1521099.



\end{document}